\documentclass[12pt,preprint]{aastex}
\shorttitle{Classical T Tauri Star BP Psc}
\shortauthors{Zuckerman et al.}

\begin{document}


\title {Gas and Dust Associated with the Strange, Isolated, Star BP 
Piscium}

\author{B.Zuckerman\altaffilmark{1,2}, C. Melis\altaffilmark{1}, Inseok 
Song\altaffilmark{3,4}, David S. Meier\altaffilmark{5,6}, \\ 
Marshall D. 
Perrin\altaffilmark{7}, Bruce Macintosh\altaffilmark{7}, Christian 
Marois\altaffilmark{8}, \\
Alycia J. Weinberger\altaffilmark{9}, Joseph H. Rhee\altaffilmark{1}, 
James R. Graham\altaffilmark{7},  \\
Joel H. Kastner\altaffilmark{10}, Patrick Palmer\altaffilmark{11}, T. 
Forveille\altaffilmark{12},   \\
E.E. Becklin\altaffilmark{1,2}, D.J. Wilner\altaffilmark{13}, T.S. 
Barman\altaffilmark{14}, G.W. Marcy\altaffilmark{7}, M.S. 
Bessell\altaffilmark{15}}
\altaffiltext{1}{Department of Physics \& Astronomy, University of 
California, Los Angeles CA 90095}
\altaffiltext{2}{UCLA Center for Astrobiology} \altaffiltext{3}{Gemini 
Observatory, 670 North A'ohoku Place, Hilo HI 96720} 
\altaffiltext{4}{present address: Spitzer Science Center, California 
Institute of Technology, Pasadena, CA 91125}
\altaffiltext{5}{David S. Meier is a Jansky Fellow of the National Radio 
Astronomy Observatory.}
\altaffiltext{6}{National Radio Astronomy Observatory, P.O. Box 0, 
Socorro, NM 87801}
\altaffiltext{7}{Department of Astronomy, University of California, 
Berkeley CA 94720}
\altaffiltext{8}{Institute of Geophysics \& Planetary Physics, Lawrence 
Livermore National Laboratory, Livermore CA 94551}
\altaffiltext{9}{Carnegie Institution of Washington, Department of 
Terrestrial Magnetism, 5241 Broad Branch Road, Washington DC 20015}
\altaffiltext{10}{Rochester Institute of Technology, 54 Lomb Memorial 
Drive, Rochester NY 14623}
\altaffiltext{11}{Department of Astronomy \& Astrophysics, University of 
Chicago, Chicago IL 60637}
\altaffiltext{12}{Laboratoire d'Astrophysique de Grenoble, Universite J. 
Fourier, B P 53, 38041, Grenoble, Cedex 9, France}

\altaffiltext{13}{Harvard-Smithsonian Center for Astrophysics, 60 Garden 
Street, Mail Stop 42, Cambridge, MA 02138}

\altaffiltext{14}{Lowell Observatory, 1400 West Mars Hill Road, Flagstaff 
AZ 86001}

\altaffiltext{15}{Research School of Astronomy and Astrophysics, 
Institute 
of Advanced Studies, Australian National University, ACT 2611, Australia}

\begin{abstract}

We have carried out a multiwavelength observational campaign demonstrating
some of the remarkable properties of the infrared-bright variable star BP
Psc.  Surrounded by a compact dusty, gaseous disk, this little-studied 
late-G (or early-K) type star emits about 75\% of its detected energy flux 
at infrared
wavelengths.  Evidence for accretion of gas in conjunction with
narrow bi-polar jets and Herbig-Haro objects is apparently consistent with
classification of BP Psc as a pre-main sequence star, as postulated in
most previous studies.  If young, then BP Psc would be one of the nearest
and oldest known classical T Tauri stars.  However, such an evolutionary
classification encounters various problems that are absent or much less
severe if BP Psc is instead a luminosity class III post-main
sequence star.  In this case, it would be the first known example of a 
first ascent giant surrounded by a massive molecular disk with 
accompanying rapid gas accretion and prominent jets and HH objects.  In 
this model, the genesis of the massive dusty gaseous disk could be a 
consequence of the  
envelopment of a low mass companion star.  Properties in the disk may  
be conducive to the current formation of planets, a gigayear or more after 
the formation of BP Psc itself.  

\end{abstract}

\keywords{stars: general --- planetary systems: protoplanetary disks: 
individual(BP Piscium)}

\section{Introduction}

As laboratories for the early stages of the formation of planetary
systems, T Tauri stars have engaged the energy of astronomers for decades.
It is desirable to find those closest to Earth because orbiting gas, dust
and protoplanets can be studied with higher linear resolution, that is,
smaller semi-major axes.  Young stars are usually found in the Galactic
plane.  As a consequence, specific effort has been expended to find T
Tauri stars at high Galactic latitudes because these promise to be
relatively close to Earth.  In addition, classical T Tauri stars are
bright infrared sources and are almost always associated with interstellar
molecular nebulosity.  Thus, early searches for nearby T Tauri stars
focused on bright IRAS sources and interstellar molecular clouds at high
Galactic latitude (de la Reza et al. 1989; Gregorio-Hetem et al. 1992;
Magnani et al. 1995).  Two decades of research have demonstrated that   
neither interstellar molecular clouds nor T Tauri stars substantially  
younger than 10 Myr are to be found closer to Earth than $\sim$100 pc.
And by 10 Myr most T Tauri stars are of the weak-lined, rather than
classical, variety.  Two well-studied, infrared-bright, nearby classical 
T Tauri stars are the apparently single TW Hya (Qi et al. 2006 and
references therein; 56 pc from Earth) and the close binary V 4046 Sgr
(Stempels \& Gahm 2004; Gunther et al. 2006; $\sim$80 pc from Earth).  

BP Psc is a high Galactic latitude (b = $-57$), bright IRAS source that 
generally has been classified as a T Tauri star.  Yet, remarkably, BP Psc 
(also known as StH$\alpha$202) has been almost completely overlooked by the 
astronomical community.  The star has been included in a few surveys, but, 
to the best of our knowledge, it has never been the focus of any 
published paper.  We present here a suite of optical, infrared and 
millimeter wavelength measurements of this extremely interesting, 
long-neglected star.  Notwithstanding that these data, some of which date 
back a decade, generally support a T Tauri star classification, overall the 
data appear more consistent with a post-main sequence evolutionary state.  
If so, then BP Psc possesses several properties never before observed in a 
first-ascent giant star.  Ultimate resolution of the mystery of the 
evolutionary state of BP Psc will require a direct measurement of its 
trigonometric parallax.

\section{Observations}

\subsection{Millimeter and centimeter wavelengths}

CO spectra and continuum emission were investigated with the 15\,m JCMT and
the 30\,m IRAM antennas, and the Owens Valley Radio Observatory (OVRO), 
SMA\footnote{The Submillimeter Array is a joint project between the
Smithsonian Astrophysical Observatory and the Academia Sinica Institute of
Astronomy and Astrophysics and is funded by the Smithsonian Institution
and the Academia Sinica.} and VLA\footnote{The Very Large Array of the  
National Radio Astronomy Observatory is a facility of the National Science
Foundation operated under cooperative agreement by Associated
Universities, Inc.} interferometers.  Single dish and SMA and OVRO
interferometer particulars appear in Tables 1 and 2, respectively. Results
from these two mm-wavelength interferometers are summarized in Table 3.
The CO spectra from four telescopes are displayed in Figure 1.  The
interferometric CO maps appear in Figure 2.

The JCMT receiver B3i was
used for measurement in the 345 GHz atmospheric window.  The backend was
the Digital Autocorrelation Spectrometer (DAS); with a bandwidth of 250
MHz, which yielded a spectral resolution of 189 kHz.  Data were obtained
in beam-switching mode.  Integration time was 95 minutes at the J = 3$-$2
CO transition.  To convert from antenna temperature to main-beam
brightness temperature we applied an aperture efficiency of 0.49.  At the
IRAM 30 m the ``front-end'' was an SiS receiver and the spectrometer was a
256-channel filter bank with 1 MHz resolution.  Data were obtained in
beam-switching mode.  Integration time was 106 minutes at the 2$-$1 CO
transition and 245 minutes for 2$-$1 $^{13}$CO, but in two polarizations 
that were then averaged together.  To convert from antenna to
main-beam brightness temperature we applied an aperture efficiency of
0.53.

Aperture synthesis observations of the $^{12}$CO(2$-$1) transition of BP Psc
were made with the Owens Valley Radio Observatory (OVRO) Millimeter
Interferometer between 1996 March 9 and 1997 January 10 (UT).  The
interferometer consisted of six 10.4 m antennas with cryogenically cooled 
SiS receivers (Padin et al. 1991, Scoville et al. 1994).  Observations in
``Equatorial'' and ``High'' configurations were obtained, with system
temperatures (single sideband) ranging between 700 and 1400 K at 230 GHz.  
One hundred twelve 0.538 MHz channels covered the transition with a 
velocity resolution of 0.70 km s$^{-1}$. 
Simultaneous wideband (1.0\,GHz) continuum observations were obtained at 3 
mm (96.627\,GHz) and 1\,mm (230.55\,GHz).  The data set was calibrated 
using the MMA software package.  The phase was calibrated by observing the 
quasar 3C454.3 every 18 minutes.  Absolute flux calibration was done using 
Neptune as primary flux calibrator and 3C454.3, 0528+134 and 3C84 as 
secondary flux calibrators.  Absolute flux calibration is estimated to be 
good to $\sim$10\% at 3\,mm and $\sim$25\% at 1 mm.  All subsequent data 
analysis and manipulation employed the NRAO AIPS package.  The OVRO 
primary beam is $\sim$32$\arcsec$ at 230 GHz.  Corrections for the primary 
beam have not been applied.  The longest spatial scales sampled are 
$\sim$11$\arcsec$ for CO(2$-$1).

We observed BP Psc with the SMA on 2006 September 26 (UT) using 7 
antennas that provided baseline lengths from 14 to 69 meters.  The 
correlator was configured for the full 2 GHz bandwidth with uniform 
channel spacing 0.8125 MHz, or 0.7 km s$^{-1}$ at the CO J=3$-$2 line 
frequency centered in the USB. Calibration of complex gains was performed 
by interleaving 4 minute observations of the quasars 3c454.3 (3.4 Jy) and 
2225-049 (1.6 Jy) with 10 minute observations of BP Psc, which was 
observed over the hour angle range $-4$ to +4.  The system temperatures 
(DSB) ranged from 175 to 400 K. The absolute flux scale was determined 
through observations of Uranus, with an estimated uncertainty $\lesssim$20\%.  
The naturally weighted CO (3$-$2) beam size was 2.4$\arcsec$ $\times$ 
2.0$\arcsec$ at P.A. 29 deg. (Table 2).  The rms noise level for the continuum 
image (Figure 3) combining the lower and upper sidebands (effective bandwidth 
$\sim$4 GHz), was 2.5 mJy. The SMA primary beam is 35$\arcsec$ FWHM at 
346 GHz. The data calibration was performed using the MIR software, 
followed by standard imaging and deconvolution using MIRIAD.

We observed BP Psc with the VLA on 28 January 2007 for a total of one 
hour (including all calibrations).  The array was in the DnC 
configuration.  The observations were at X-band (on source 20 minutes) 
and L-band (on source 13 minutes) with the default settings for 50 MHz 
bandwidth. At both frequencies the phase calibration source was 2330+110 
and the absolute flux calibrator was 0137+331 (3C48).  Nominally, 27 
antennas were in use; but two did not provide any useful data.  The 
operator noted intense RFI most of the time at L-band.  The X-band image 
covers an area of 10.2$\arcmin$ $\times$ 10.2$\arcmin$, the final L-band 
image covers an area 51.2$\arcmin$ $\times$ 51.2$\arcmin$

\subsection {Infrared}

Spectra, photometry and adaptive optics imaging were obtained in the mid- 
and near-IR with instruments on the Keck and Gemini telescopes at Mauna 
Kea Observatory.  BP Psc was observed with the Keck Observatory 
adaptive optics (AO) system 
(Wizinowich et al. 2006) and the facility IR camera NIRC2 on 19 July 2006 
(UT).   At $J, H, K'$ and $L'$ wavebands we obtained 4 images, each 
consisting of 50 coadds with 0.2 second exposure time. We also observed a 
nearby point spread function (PSF) reference star of similar brightness. 
The reference star images have a full-width-half-maximum (FWHM) of 0.050 
arcseconds at K with a Strehl ratio of 0.3; the H images have a FWHM of 
0.045 arcseconds and a Strehl ratio of 0.18. The PSF reference star is 
clearly pointlike, indicating the elongation of BP Psc seen in Figure 4 
is not an AO artifact. We used Lucy-Richardson deconvolution to enhance 
the resolution of the BP Psc images.  Figure 4 shows H and K' band images 
of BP Psc together with the PSF reference star and the deconvolved 
images. 

Gemini/MICHELLE mid-infrared photometry and spectra of BP Psc were 
obtained on 2006 October 4-5 (UT) (Table 4 and Figs. 5 \& 6).  The water 
vapor (PWV) during the 4 October observations was 5 - 6 mm, as estimated 
from converting the 225 GHz optical depth measured at the Caltech 
Submillimeter Observatory to precipitable water vapor according to Davis 
et al (1997). Consequently, there was a high and probably variable 
infrared background and lower than usual sensitivity. PWV on 5 October 
was approximately 2.8 mm.  A standard chop-nod scheme was employed; only 
the main beam was used in analysis because the telescope does not guide 
in the sky beams.   Observations of BP Psc were interspersed with those 
of photometric and telluric spectral standards at similar airmass. 
Because of the large PWV on 4 October, the standard star flux in the 
MICHELLE filters was estimated by taking the Cohen et al. (1999) template 
spectrum and multiplying it both by the filter transmission and an ATRAN 
model of 5 mm PWV. The Si-1 filter is affected by loss of transmission at 
wavelengths $<$8 $\mu$m through the atmosphere at this high PWV, 
but the other N-band 
filters are largely unaffected.  Quoted flux uncertainties are larger 
then the statistical uncertainties and were estimated from variability in 
the standard stars where possible as well as from experience with mid-IR 
observations from Mauna Kea.

For the low spectral resolution (R$\sim$200) N-band spectra, wavelength 
calibration was performed by cross-correlating the telluric ozone feature 
in both spectra with the high fidelity spectrum from the National Solar 
Observatory (Wallace et al. 1994). A small offset (0.2 pixel = 0.004 
$\arcsec$) in wavelength between the BP Psc and standard spectra produced 
the lowest noise in the Ozone region of the divided spectrum. The measured 
dispersion was 0.02587 $\mu$m/pixel. The final N-band spectrum was 
normalized to the flux density in the Si-5 filter, which is completely 
contained within it and in a clear part of the atmospheric window.

Aperture photometry of BP Psc is reported in Table 4. The N-band psf
varied in size during the course of the observations. To eliminate any 
uncertainty associated with aperture corrections, a large, 12-pixel (1.5 
$\arcsec$ radius) aperture was used for photometry.  Uncertainties were 
estimated from the standard deviation of sky pixels in an annulus at 
15-20 pixels. In the case where more than one observation was made at a 
given wavelength or spectral mode, the results were averaged.  The 
spectra were extracted by summing 9 pixels, approximately twice the FWHM 
in the spatial direction, centered on the source.  Uncertainties at every 
wavelength were estimated from the standard deviation of sky pixels in a 
1$\arcsec$ region 1$\arcsec$ away from the spectrum.

Near-IR spectra were obtained on 09 August 2006 UT with NIRSPEC (McLean et 
al. 1998, 2000) on the Keck II telescope (Table 5).  In addition, a 
spectrum in high-resolution mode in the K-band window was obtained on 22 
July 2006 UT and in low-resolution mode in the J-band on 04 August 2006 
UT.

\subsection {Optical}

We obtained Keck HIRES echelle (Vogt et al. 1994) spectra of BP Psc at 
five epochs (Table 5 and Figs. 7$-$9).  Wide field images (Figures 10 \& 
11) centered on BP Psc were acquired on 23 Oct. 2006 (UT) with a set of 
large (4-inch) high-efficiency H$\alpha$ and [S II] filters recently 
obtained by J. R. Graham and M. D. Perrin for use with the prime focus 
camera (PFCAM) on the 3 m Shane telescope at Lick Observatory.  Total 
exposure time with each filter was 40 minutes, but these were mosaicked so 
not all pixels reached the same sensitivity.  In addition, a 420 sec 
exposure was obtained in the continuum with a Spinrad R-band filter.  The 
seeing was typical for Lick and the measured FWHM of the PSF is 1.67 
$\pm$0.07\arcsec.  Subsequently, an image of the NE jet was obtained under 
better seeing conditions (FWHM $\sim$0.65$\arcsec$ measured from point 
sources), with the GMOS camera and g-band, i-band, and H$\alpha$ filters 
at the Gemini North Observatory (Fig. 12); exposure times in the three 
filters were 210, 210, and 1800 sec., respectively.

A low dispersion spectrum was taken on May 25 2007 UT 18:10 with the ANU 
Double Beam Spectrograph (DBS) on the 2.3 m telescope at Siding Spring 
Observatory. The resolution in the blue, between 3300 and 5200 \AA\ was 
about 3 \AA\ and about 5 \AA\ in the red between 6300 and 10000 \AA.  The 
overall shape of the DBS spectrum agrees well with one obtained with the 3.6 
m ESO telescope at La Silla Observatory (Suarez et al 2006 and O. Suarez 
private communication).  The DBS blue spectrum (Fig. 13) showed strong 
emission in the Ca H and K lines as well as emission at 4068 \AA.

\section {Results}

Perhaps the earliest indication that BP Psc is an unusual star was the
1996 single antenna radio observations of Table 1 and Figure 1.  Very few
nearby, high Galactic latitude, isolated, main sequence and pre-main
sequence stars have detectable CO pure rotational emission; TW Hya and 49 
Cet are probably the most noteworthy (e.g., Zuckerman et
al. 1995a).  No first-ascent giant star is known to display such radio
emission. The double-peaked single dish spectra showed that probably the CO
at BP Psc is distributed in an orbiting ring.  A non-detection of the
$^{13}$CO J = 2$-$1 transition with the IRAM 30\,m telescope at a level 7
times fainter than the $^{12}$CO 2$-$1 line implies a $^{12}$CO opacity
$<$14 if the isotope ratios are solar (1/89).  This result indicates that
the CO lines are not very optically thick and that the molecular disk 
is substantially less massive than disks orbiting some young T
Tauri stars located near molecular clouds (see Section 4.7).

The single dish spectra can be compared with the SMA and OVRO spectra to 
ascertain whether the interferometers are detecting all of the CO 
emission.  Figure 1 shows this indeed to be the case, and the right hand 
column of Table 2 indicates that the interferometers detect essentially 
all the flux seen with the 30 m and JCMT.  The SMA spatial resolution was 
(barely) sufficient to resolve the region of CO 3-2 line emission along a 
position angle consistent with that of the dust disk implied by our Keck 
AO images (Figure 4 and below).  The SMA detection of 0.88 mm continuum 
emission (Table 3 and Figure 3) enables an estimation of the mass in dust 
particles orbiting BP Psc (Section 4.6).  Given the presence of bipolar 
jets emanating from BP Psc, radio continuum emission might be detectable 
(e.g., Reipurth \& Bally 2001), but the results of our VLA observations 
were negative (Section 4.9).

Turning to the infrared and optical, our Keck AO images show a curved 
morphology with no central point source consistent with a near-edge-on 
disk optically thick at near-IR wavelengths (Figure 4).  The 
sky-projected position angle of the major axis of the disk lies 
(somewhere) in the range between 113 and 123$^{\circ}$, east of north.  
Comparison with disk modeling in Burrows et al. (1996) and Wood et al. 
(1998) suggests a disk inclination of 70$-$80$^\circ$ (0$^\circ$ is face-on).  
The earth-facing polar axis of the disk lies at a position angle in the 
range 203 to 213$^\circ$.  That is, the side of the disk tilted slightly 
so that it comes into view along our line of sight is to the SW.  In the 
deconvolved images, some hint of the side of the disk that is tipped away 
from us (to the NE) is visible, the classic morphology of a near-edge-on 
T Tauri disk (e.g. Burrows et al. 1996).  However, the BP Psc disk, of 
visible extent $<$0.2\arcsec, is much more compact than those seen in Taurus.  
The AO images indicate that only light scattered toward Earth by a 
portion of the disk is seen in the JHK filters, but not the star itself.  
Consequently, we can be quite sure that the star is not seen directly at 
any optical wavelength.

Figure 6 displays the spectral energy distribution (SED) of BP Psc.  To
derive the total energy flux received at Earth, we combine the flux from a
5000 K blackbody with that from two colder blackbodies at 1500 and 210 K to
represent the infrared emission from the disk.  While this model for the
SED is not unique, it produces a reasonable match to the data and implies
the presence of very hot dust, consistent with the evidence for J- and
H-band excesses in classical T Tauri stars presented by Cieza et al. (2005)
and references therein.  The flux decomposition indicates excess emission
from BP Psc even in the J band.  Overall, $\sim$75\% of the bolometric 
luminosity
of the star, as seen from Earth, is reprocessed by the orbiting dust grains
and re-emitted at IR wavelengths.  Such a high percentage is unprecedented
for a star at high Galactic latitude far from any interstellar molecular
cloud (e.g., Figure 2 in Zuckerman et al. 1995a).  In addition, the SED of
BP Psc implies the existence of much more warm circumstellar dust than is
present around the nearby classical T Tauri stars TW Hya and V 4046 Sgr.

If BP Psc is a pre-main sequence star, the total luminosity of the system,
at each age given in Table 6, can be calculated with the listed radii and
a temperature of 5000 K.  The corresponding distances (D) from Earth
listed in the 4th column assume: (1) all of the detected energy under the
SED is generated at the photosphere of BP Psc and (2) the flux emitted in
our direction would not be substantially different if BP Psc were to be
observed from a very different direction (e.g., at inclination 0$^\circ$).  
However, as some of the observed luminosity can be generated by mass
accretion onto BP Psc, and because of the large mid-plane optical
extinction, both assumptions are questionable.  When these two effects are
estimated and included, revised distances (D$'$) at each young age are
listed in the right hand column of Table 6.  Details, and consideration of
the distance to BP Psc if it is either a pre- or post-main sequence star,
are presented in Section 4.3.

The Gemini spectrum (Figure 5) and photometry are consistent with earlier
IRAS photometry and, importantly, show little or no indication of any
silicate features in the 10 $\mu$m window.  The wavelength region of 9.3-10
$\mu$m includes strong telluric ozone absorption that varies with time and
airmass.  Although division by the standard star mainly removes its 
effect,
albeit with larger uncertainties at these wavelengths, some residuals
appear as wiggles in the spectrum where the slopes of the absorption lines
are large.

With NIRSPEC, emission lines are seen from H I Paschen $\beta$, Brackett
$\gamma$, and Pfund $\delta$, and, also, near 1.257 ([Fe II]) and 1.2944
microns (probably due to [Fe II]).  Many additional emission lines, both
permitted and forbidden, are seen in our HIRES data.  The strong, broad
H$\alpha$ emission line is consistent with classification of BP Psc as a
classical T Tauri star (Section 4.3).  Spectra of young stars with
microjets and Herbig-Haro objects, of the sort seen in Fig. 7$-$9, usually
display optical forbidden lines.  If a star is surrounded by a dusty disk,
viewed close to edge-on, as per BP Psc, then the forbidden lines are
usually blue shifted with respect to the systemic velocity.  We interpret
the central (heliocentric) velocity of the CO J = 3-2 line measured with
the SMA, $-17.6$ km s$^{-1}$ (Table 3) to be the velocity of BP Psc.  The
radial velocities of the [OI], [NII], and [SII] lines displayed in Figs.  
8 and 9 are blueshifted, typically by about 8 km s$^{-1}$ from this
systemic velocity.  Although the radial velocity of the photospheric
absorption lines is variable (Table 5 and Section 4.5), the causes of the
variations are still uncertain.

In addition to emission lines from H, Ca, O, N, Fe and S, emission lines
of neutral helium also appear to be present.  A weak He I 5877.2 \AA\
(vacuum) line appears in all five epoch HIRES spectra with about the same
intensity and line shape.  By contrast, a more prominent He I emission
line appears at 8446.8 \AA\ in both the 01 and 09 Sept. 2006 spectra, but
with very different line shapes.  A similar striking variation is evident
in the shape of the nearby Ca II infrared triplet line at 8500.4 \AA.

With moderate resolution (3.5 \AA) spectra, Whitelock et al (1995)  
reported the presence in absorption of the G-band of CH near 4300 \AA\ and
CN bands around 3885 and 4215 \AA.  Our HIRES spectra show the G-band and
perhaps the CN bands, while the DBS spectrum (Fig. 13) reveals both the CH
and CN bands.  The MgI b triplet near 5175 \AA\ is evident in absorption
in the HIRES spectra.

Our wide field optical images (Fig. 10 - 12) reveal a chain of Herbig-Haro
objects extending over a total of $\sim$10$\arcmin$ (Table 7).  If BP Psc
is a pre-main sequence star at a distance of 100 pc, then the HH objects
are seen over a total linear span $>$0.3 pc, and over an appropriately
greater span if BP Psc is a giant star at a few hundred pc.  The bipolar
outflow extends to the NE (position angle $\sim$24$^\circ$) and southwest
of BP Psc, consistent with the orientation of the polar axis of the
circumstellar disk seen in our Keck AO and SMA images.  Based on the
disk's apparent inclination (see above), we anticipate that the NE jet is
redshifted and the SW jet blueshifted.  Details of the outflow are
discussed in Section 4.9.

\section {Discussion}

\subsection {Spectral type}

As noted in Section 3, BP Psc is not directly visible at any optical or 
near-IR wavelength. Thus, the SED cannot be used to deduce a 
spectral-type.  Perhaps this accounts for the varied spectral types 
that have appeared in the published literature (mid-F, Downes 
\& Keyes 1988; Ae, Gregorio-Hetem et al. 1992; G?, Whitelock et al. 
1995).  To deduce the spectral type, we compared our HIRES spectra with 
the spectra of a variety of main sequence stars without surrounding dust 
(see Figure 14).  If a dwarf, then the spectral type of BP Psc is early-K; 
we will adopt K2 and T$_{eff}$ = 5000 K.   If BP Psc is a luminosity class
III giant star, then, at about this temperature, its spectral type is late-G 
(see Section 4.3.2).

While the SED cannot reliably be used to establish the spectral type of BP 
Psc, the following result from VizieR is noted.  VizieR lists measurements 
of optical broadband colors from a wide variety of sources, obtained at 
various epochs; nine at V-band range over 1.3 magnitudes (11.53$-$12.85) 
and 5 at R-band range over 0.5 mags (11.0$-$11.5).  The range in 
brightness is due to intrinsic variability of this known variable star 
combined with some measurement error.  An average of these brightnesses 
yields a V$-$R consistent with an early K-type dwarf or late G-type giant 
(V$-$R $\sim$0.7 mag) with little or no reddening, consistent with the 
conclusion we draw from our DBS spectrum (Section 4.3.2).

\subsection {Proper motion}

If the distance to BP Psc were known, then we could decide if it is a 
giant or a dwarf.  But no trigonometric parallax has been measured and, 
instead, we use proper motion to help estimate a distance to BP Psc.  The 
Tycho 2 proper motion of BP Psc is 44.4 $\pm$ 4.1 and $-26.3$ $\pm$ 4.3 
mas yr$^{-1}$ in R.A. and Decl., respectively. Because of its unusual spectrum 
(e.g., Fig. 7) and variable radial velocities (Table 5 and Section 4.5), 
plausibly, BP Psc might be a close binary star; these often presented 
problems in measurement of Hipparcos/Tycho parallaxes and proper 
motions.  Also, as noted in Section 3, Tycho was measuring the position 
of light scattered off dust orbiting BP Psc, not direct starlight 
itself.  Thus the Tycho proper motion measurement could be adversely 
affected.  However, the proper motion error given above is typical of 
errors in proper motion of single stars of brightness comparable to BP 
Psc measured by Tycho.  In addition, the Lick Northern Proper Motion 
Program: NPM1 Catalog (Klemola et al. 1987) gives a proper motion of 36.6 
$\pm$ 5, $-42.6$ $\pm$ 5 mas yr$^{-1}$, in reasonable agreement with Tycho, in 
particular, in the important absolute value of the total proper motion. 

We can also use the epoch 2006.7 SMA measurements of the dust emission
centroid (Fig. 2) and the epoch 1991.7 Tycho-2 position to check the
cataloged Tycho-2 proper motion.  The epoch 1991.75, equinox J2000, Tycho-2
position is 23h22m24.6666s, $-02$d13\arcmin41.049\arcsec.  The total 15
year proper motion, based on the Tycho-2 proper motion (see above) is 666
mas in R.A. and $-394$ mas in Decl., indicating a J2000, 2006.7 epoch,
position of 23h22m24.711s, $-02$d13'41.44".  The errors in this position
are 150 mas and 188 mas respectively, in R.A. and Decl., where the errors
are derived by adding, in quadrature, the Tycho-2 errors in the 1991.7
position and the proper motion.  Within the errors, this Tycho-2 predicted
2006.7 position for BP Psc is in agreement with the SMA measured position
(see caption to Fig 2 and the 880 $\mu$m entry in Table 3).  In sum, there
is no reason to doubt the accuracy of the Tycho 2 proper motion.

\subsection {Distance, bolometric luminosity, and evolutionary state}

In the literature, BP Psc has usually been classified as a T Tauri star.  
However, such an interpretation is not without its problems as described
below.  So, in the absence of a direct trigonometric measurement of the
distance from Earth of BP Psc, it is prudent to consider alternative
evolutionary states.

Some post asymptotic giant branch (AGB) stars are accompanied by 
substantial quantities of dust and molecular gas, with a substantial 
fraction residing in orbiting circumstellar disks (e.g., de Ruyter et al. 
2006).  Post-AGB stars have bolometric luminosities a few 1000 times 
solar or more.  Integrating under the SED (Fig. 6), if BP Psc has a 
luminosity of 3000 L(sun), then it is 5400 pc from Earth.  At a distance 
of 5400 pc, the Tycho 2 proper motion implies a velocity in the plane of 
the sky of 1400 km s$^{-1}$.  Clearly, BP Psc is not a post-AGB star.  

\subsubsection {Classical T Tauri star?}

If BP Psc is a pre-main sequence star, then its distance from Earth and
Galactic space motion (UVW) depend on its age.  Table 6 lists some
age-dependent properties based on evolutionary models of Baraffe et al.
(2003 and private comm. 2006).  Table 6 UVWs are similar to the motion of
stars in the solar vicinity with ages between about 10 and 100 Myr (Table
7 in Zuckerman \& Song 2004 and Song et al. 2007).  While not a proof of
youth, the fact that, as a dwarf, BP Psc turns out to have the space
motion of a young star, rather than some completely different motion,
gives weight to a young star interpretation.  In addition, the rapid
rotation of BP Psc determined from the HIRES spectra ($v$sini = 32.3
$\pm$ 1.2 km s$^{-1}$) is consistent with a very young age.  However, as
noted in Section 3, essentially all the light measured by HIRES has been
scattered off the surrounding dust and, hence, Doppler broadened.  Thus,
the measured $v$sini should be regarded only as an upper limit to the
true $v$sini of the star.

The age dependent distances to BP Psc, column 4 in Table 6, were
derived in the following way.  First we fixed the temperature at 5000
K from the HIRES echelle spectrum (Section 4.1).  Then we chose an age
and found the appropriate stellar radius from the Baraffe et al.
models.  With this radius and temperature we fitted the optical
portion of the SED (Fig. 6) with a 5000 K photospheric model and
derived a ``5000\,K distance'' to BP Psc under the assumption that
this photospheric emission encompassed all the flux.  Obviously, this
assumption is wrong because the 5000 K model accounts for only about
25\% of the total flux under the SED.  We thus corrected the derived
5000 K distances by bringing BP Psc a factor of two closer to Earth
and it is these corrected distances, D, that appear in column 4 
of Table 6.

However, two additional phenomena must also be considered.  First, as
noted in Section 4.8, perhaps 20\% of the bolometric luminosity of the
system is due to accretion of disk material onto BP Psc.  If so, then
the distances listed in Table 6 for each Baraffe model would need to
be increased by 10\%.  Given various model uncertainties, this small
distance correction, by itself, is ``lost in the noise''.

A potentially more significant correction to Table 6 distances arises
from the location of our viewing angle nearly in the plane of BP Psc's
dusty disk.  The energy contained under the 210 K black body curve in
Fig. 6 is roughly half the energy under the sum of the 5000 K and 1500
K curves.  The former is thermal emission from the grains and should
be essentially constant independent of viewing angle (because optical
depths at these long wavelengths are small).  But the shorter
wavelength optical plus near-IR light can vary substantially with
viewing orientation.  Estimates for face-on disks can be obtained from
Weinberger et al (2002) or Low et al. (2005), showing the SED of TW
Hya, and from Figure 1b in Wood et al. (1998).  From these SEDs, we
estimate that, for face-on orientations, the short wavelength flux
from BP Psc could be about 7 times the long wavelength thermal
emission (rather than only twice as large, as per Fig. 6).  If
observed from the same Earth-BP Psc separation but in the face-on disk
direction, the total flux received from BP Psc would be about 3 times
larger than the flux indicated in Fig. 6.  Then, if averaged over all
4$\pi$ steradians, the total energy emitted by BP Psc might be about
twice what one would deduce from assuming received flux is independent
of viewing orientation and equal to that in Fig. 6.  If so, then the
distances D listed in Table 6 all should be reduced by a factor of
1.4.

The above two correction factors operate in opposite directions and
each is obviously uncertain.  Nonetheless, the distances D to BP Psc
listed in Table 6 may all be overestimated by a factor of about 1.3.  
For example, if the Baraffe et al. models are accurate, then if 10 Myr
old, BP Psc would be only 80 pc from Earth.  Actually, for the Tycho 2
proper motion of 52 mas yr$^{-1}$, 80 pc would be a more typical
young-star distance than 104 pc.  In view of the above discussion, in
the right hand column of Table 6, for each age we list distances D$'$
that are a factor of 1.3 times smaller than the distances D listed in
the fourth column.

Given the existence of massive amounts of dust (Section 4.6), an
orbiting gaseous disk (Section 4.7), and prominent polar jets and
Herbig-Haro objects (Section 4.9), if BP Psc is a pre-main sequence
star, then we tentatively adopt an age of $\sim$10 Myr; a still
younger age would compound both the weak lithium line problem and the
mystery of BP Psc's origin (Section 4.10).  The measured equivalent
width (EW) of the lithium 6709.6 \AA\ line is only 50 m\AA, (Figs. 9
\& 15) which is more appropriate for a K-star of age 100 Myr or
greater (Figure 3 in Zuckerman \& Song 2004).  We know of no other
example of an early K-type star of age 10 Myr with such a weak lithium
line.  Because significant mass is accreting onto BP Psc (Section
4.8), one should consider the possibility of severe veiling at the Li
line, even though veiling at 6700 \AA\ in an early K-type star is
usually modest compared to veiling at other wavelengths in typical T
Tauri stars (see, e.g., Figure 5 in White \& Hillenbrand 2004).  
Following the method of Palla et al.  (2005) we compared the strength
of the 6678 and 6710 \AA\ lines of Fe I and the 6768 \AA\ line of Ni I
in BP Psc and in 7 main sequence dwarfs with V-K colors of early
K-type stars, but without veiling.  In this way we estimate a
veiling of about 15\% and, at the most, veiling reduces the Li line EW at 
BP Psc by about 20 m\AA.  
But even at 70 m\AA, the Li line EW in BP Psc would be characteristic
of early K-type stars 100 Myr old or older.

Considering the characteristic long time, hundreds of millions of
years, for substantial lithium depletion in early K-type stars, such a
weak line is not easily understood if BP Psc is only $\sim$10 Myr old.  
Other heavily dust enshrouded, isolated, K-type classical T Tauri
stars (e.g. TW Hya [Webb et al. 1999] and V4046 Sgr [Stempels \& Gahm
2004]) have lithium absorption lines with EW $\sim$500 m\AA .  Indeed
V4046 Sgr is a close binary and both K-type components show these very
strong lithium lines.

If BP Psc is 10 Myr old, then it may be the oldest known pre-main
sequence or main sequence star associated with bipolar jets and HH
objects (e.g., Reipurth \& Bally 2001).  At the corresponding distance
of 100 pc or less (Table 6), BP Psc would be the closest known T Tauri
star with spatially resolved polar jets (plus HH chains).  If it is a 
pre-main sequence star, then whatever its precise age,
BP Psc clearly should be classified as a classical
T Tauri star based on the definition of White \& Basri (2003, see also
Briceno et al. 2007): a star with spectral type in the range K0 to K5
with H$\alpha$ emission equivalent width $>$3 \AA\ and full width at
10\% of peak flux $>$270 km s$^{-1}$.  Our three measurements of EW
(Table 5) are all much larger than 3 \AA\ (see also Gregorio-Hetem et
al. 1992) and the 10\% width of the H$\alpha$ line (Figure 9) is
$\sim$500 km s$^{-1}$.

\subsubsection {First ascent giant or subgiant star?}

If BP Psc is a first ascent giant star, then it is the first one known 
with a gaseous disk detected in pure rotational CO transitions and 
undergoing rapid accretion onto the central star.  It would also be the 
first known with bipolar outflow and associated Herbig-Haro objects.  

Notwithstanding these unprecedented properties, the data for BP Psc better
support first-ascent giant status than any alternative evolutionary state.  
The lithium line EW is nothing special for a first ascent giant star;  
depending on distance, a good match can be made between the kinematic mass
and evolutionary mass (see Section 4.4); and the Galactic space motions
are entirely reasonable for a (few) Gyr old star.  For example, at a
distance of 300 pc (see Section 4.4), UVW = -38.8, -61.0, -21.9 km
s$^{-1}$.  The important V component is typical of white dwarfs (e.g.,
Table 6 in Zuckerman et al. 2003) having ages that are likely to be
comparable to that of BP Psc if it has evolved from an A- or F-type main
sequence progenitor.  Perhaps the most telling reason why BP Psc is
probably a first ascent giant star are the relative intensities of 
gravity sensitive lines, as outlined in the following discussion. 

Fekel et al (1996) used luminosity-sensitive line ratios in the vicinity of 
6440 \AA\ to help them deduce that HD 233517 is a dusty, lithium-rich, first 
ascent, giant star (rather than pre-main sequence).  To investigate the 
luminosity class of BP Psc, we first used iron and calcium lines in our 
HIRES spectra in this spectral region (Fig. 16).  This comparison 
suggested that BP Psc is probably a giant star or perhaps a subgiant, but 
probably not a dwarf.  Given this result and to better pursue this path, we 
obtained the DBS spectrum mentioned in Section 2.3 and compared it to a 
variety of spectral standards.

Except for much stronger absorption in H$\delta$ and H$\gamma$, the 
underlying DBS absorption line spectrum resembles that of G8III stars in the 
MILES database (Sanchez-Blasquez et al 2006).  The blue spectrum of BP Psc 
does not resemble late-G or early-K dwarfs because it shows weaker neutral 
lines of Fe and Mg and strong bands of CN at 4115 and 3889. In Fig. 13 the 
relative F$_{\lambda}$ spectrum of BP Psc is shown in comparison with one 
of 
these G8III stars HD 216131 from the Miles database smoothed to match BP 
Psc.  As well as the line spectrum being a good match, the relative fluxes 
are in good agreement indicating that BP Psc has little reddening.

The ELODIE Archive (Moultaka et al 2004) of echelle spectra also has many 
late-G giants and dwarfs for comparison with the HIRES echelle spectrum of 
BP Psc. We initially selected a region near 5350 \AA\ that contained several 
well defined FeII and TiII lines in addition to other lines of neutral 
species (Fig. 17). The Elodie spectra were again broadened to match the 
rotationally broadened lines of BP Psc. The comparison indicated clearly 
that the spectrum of BP Psc resembled that of a luminosity III star and not 
a class V or II star. Furthermore, the strongest features in BP Psc were 
systematically shallower than the same features in the comparison stars, a 
systematic effect that was eliminated by adopting a 15\% veiling of the BP 
Psc spectrum.  Comparison of BP Psc with the same three standard stars, but 
near 6060 \AA, yields similar conclusions (Fig. 18).

The region of the H$\delta$ and H$\gamma$ lines were also compared in the 
echelle spectra in order to assess their unusual strengthing in BP Psc. A 
value of 15\% veiling in the blue was also indicated and anything larger 
would put the core of H$\delta$ below zero. The hydrogen lines in BP Psc 
had 
almost twice the equivalent width of the comparison G8III stars and stronger 
even than in a much hotter F8III star. In addition, the shape of these two 
hydrogen lines was quite different from either a G8III or F8III star having 
deeper and wider cores and shallow wings.

Independently, Katia Biazzo and Antonio Frasca at Catania Observatory used 
their ROTFIT code (Frasca et al 2003 \& 2006; Biazzo et al 2007) to compare 
several different red wavelength regions in the BP Psc Keck spectrum with 
stars in the Elodie Archive. They derive a vsini of about 38 km s$^{-1}$ 
for BP Psc, deduce that the spectral-type best reproducing these regions 
was G8-K0 III/IV, and rule out class V dwarfs. Furthermore, a veiling of 
15\% improved the residual fits and strengthened the giant luminosity 
identification. They suggest a $T_{eff}$ $\sim$4900$\pm$100 K and logg 
$\sim$2.5$\pm$0.3.

Although the best fit of surface gravity sensitive lines demonstrates that
BP Psc is a giant instead of a main sequence star, because young pre-main
sequence stars also have lower surface gravities than do main sequence
stars, one has to check whether a really young star model spectrum of BP 
Psc can reproduce the above-quoted logg value ($\leq$2.8).

Based on theoretical evolutionary models of Baraffe et al. (1998) and Siess 
et al. (2000), very young ($\sim$1 Myr) pre-main sequence stars with 
effective temperature similar to that of BP Psc ($\sim$5,000 K) have logg 
$\sim$3.6-3.7 which is significantly higher than the best fit value. This 
strongly supports classification of BP Psc as a post-main sequence giant 
of luminosity class III instead of as a young, pre-main sequence star.

\subsection {Kinematic mass vs. evolutionary mass}

Assuming Keplerian revolution, the spatial separation of the blue and red 
peaks in the SMA CO 3-2 emission map (Fig. 2) can be used to estimate 
the 
mass of BP Psc (e.g, Beckwith \& Sargent 1993).  Neglecting the disk mass 
(Sections 4.6 \& 4.7) compared to the stellar mass (M$_{*}$), one can 
write: 
\begin{equation} M_* = R(V_d)^2/G = R(V_p)^2/G(sini)^2 
\end{equation} 
where R is the CO disk radius, V$_d$ is the Keplerian velocity at R, V$_p$ 
is the projected velocity along the line of sight, and $i$ is the 
inclination angle of the disk to the line of sight ($i$ = 0 is face-on).  
From the adaptive optics images, we estimate the inclination angle to be
75 $\pm$ 10 deg.  The separation of the red and blue peaks is 1.2 $\pm$
0.20\arcsec (statistical uncertainty only).

A difficult issue is the value to assign to V$_p$.  According to the 
Beckwith \& Sargent (1993) models, a disk with a sharp outer edge and 
inclination angle of 75 deg should display two CO emission peaks with a 
deep central depression (their Fig. 5).  As may be seen in Fig. 1, the 
observed spectral central dip is usually weak or not even present (SMA, 
3-2 line). Examination of the interferometer maps and model fits to the 
line profiles together suggest that V$_p$ lies somewhere in the range 3 
$-$ 4.5 km s$^{-1}$, with the lower values preferred.  
For a 10 Myr old star, V$_p$ 
= 3 km s$^{-1}$, and the Table 6 dust-corrected distance (D$'$) of 80 pc, 
the derived (kinematic) value of M$_{*}$ is 0.52 M$_{\odot}$.  If, 
rather, V$_p$ = 4 km s$^{-1}$, then M$_{*}$ = 0.92 M$_{\odot}$. 

The evolutionary model-mass of a 5000 K, 10 Myr old star is 1.26 
M$_{\odot}$ (Table
6; Baraffe et al. 2003, I. Baraffe, private comm. 2006).  For the Baraffe
models the ratio of kinematic to evolutionary mass hardly varies for ages
between 3 and 20 Myr:
\begin{equation}
M_{kin}/M_{evol} \varpropto R/D^x \varpropto D/D^x  
\end{equation}
where, from columns 2 and 4 in Table 6,
the distance dependence of the evolutionary mass goes like D$^x$, and $x$
is of order unity or slightly less.  Hence, the Baraffe model mass 
appears
to be larger than the most likely kinematic mass of a pre-main sequence
star.  The discrepancy would be greater yet should BP Psc be a
single-lined spectroscopic binary with a faint companion (Section 4.5).
However, as the kinematic mass is derived from an SMA map in which the CO
emission region is barely resolved, a more definitive mass comparison
should be based on a higher spatial resolution interferometric image (as
is possible using longer baselines with the SMA).

With post-main sequence evolutionary tracks from Schaller et al (1992) or 
Girardi et al (2000), agreement between kinematic and evolutionary 
mass can be obtained if BP Psc is a giant star of mass about 1.8 
M$_{\odot}$ at a distance from Earth of about 300 pc.  If so, then the 
main-sequence progenitor would likely have been of late-A or early-F type, 
if BP Psc is a single star.  The progenitor might have been of somewhat 
later spectral type if BP Psc is a spectroscopic binary (Section 4.5), or 
if it has already accreted a companion star (Section 4.10.2).

\subsection  {Is BP Psc a spectroscopic binary star?}

The radial velocities listed in Table 5 are based on cross correlation with 
radial velocity standards.  For the three 2006 epoch spectra, we used the 
atmospheric B-band (6860-6890 \AA) to ensure proper wavelength corrections 
and to correct time variations of the zero-point wavelength shift due to 
changing observing conditions (temperature and pressure) over many nights.  
Thus, the radial velocity errors given in Table 5 for these 2006 epoch 
spectra include all uncertainties of which we are aware.  In the 1996 epoch 
spectra of BP Psc, telluric lines near 5921 \AA\ could be identified, but 
not in the 1996 epoch radial velocity standards for which any telluric 
lines were lost in a forest of many photospheric lines.  
Therefore, using the above telluric lines from the two epochs, 
the two 1996 epoch BP Psc spectra were shifted into the 01 Sep. 
2006 spectrum's 
rest-frame.  They were then cross-correlated with the 01 Sep. spectrum to 
obtain radial velocities.  All errors are included in the 1996 epoch radial 
velocity measurements listed in Table 5, including the 01 Sep. 
velocity error added 
in quadrature with cross-correlation errors. Relative to the radial velocity 
on 1 Sep., the velocity on 12 July 1996 was more positive by 0.78$\pm$1.52 
km s$^{-1}$, while the 10 Oct. 1996 velocity was more positive by 
4.06$\pm$2.00 km s$^{-1}$.

Variable Stars One-shot Project (Dall et al. 2007) observed BP Psc on June 
30 (UT) and July 14 (UT) 2006 and high precision radial 
velocities\footnote{Based on data provided by the VSOP collaboration, 
through the VSOP wiki database operated at ESO Chile and ESO Garching.} from 
these data are $-$12.9 and $-$13.5 km s$^{-1}$ with a conservative one sigma 
uncertainty of 10 m s$^{-1}$.  In addition, Torres et al. (2006) observed BP 
Psc 11 times and their measured mean radial velocity is $-$5.8 $\pm$ 2.0 km 
s$^{-1}$.  They flagged BP Psc as SB1 and classified it as G9IIIe.  Dall et 
al. also note a hint of a secondary in their cross correlation 
function (T. Dall, 
priv. comm.).  Therefore, BP Psc may be a binary star.

The range of radial velocities reported by Torres et al (2006), Dall et al 
(2007) and in Table 5 covers $\sim$10 km s$^{-1}$.  If we assume the mass of 
BP Psc is 1.5 M$_{\odot}$ and if it is orbited by an 0.1 M$_{\odot}$ 
companion at a distance of 10$^{12}$ cm, then BP Psc's projected orbital 
velocity would be $\sim$10 km s$^{-1}$.  Thus, such a possibility is 
consistent with currently available radial velocity data.
If BP Psc is a pre-main sequence star, the existence of an unseen M-type 
secondary of 0.1 M$_{\odot}$ (or greater) would widen the discrepancy 
between the estimated kinematic and evolutionary masses (Section 4.4).  

At this juncture, it seems prudent to retain the possibility that the 
modest secular variations in the measured radial velocities are due to 
scattering off dust particles in a revolving inhomogeneous envelope (in 
preference to orbital motion of a binary star).  Based on the CO line 
profiles, $\sim$10 AU from the star (as per the dimensions of the 
scattered 
light AO image in Fig. 4), a characteristic dust orbital velocity is 
$\sim$10 km s$^{-1}$.  Understanding the cause(s) of the velocity 
variations will require a comprehensive monitoring campaign.

\subsection {Dust mass}

The SMA detected an 880 $\mu$m continuum flux density, S$_{\nu}$ of 18 
mJy (Table 3; Fig. 3).  We assume S$_{\nu}$ is generated entirely by dust 
particles in orbit around BP Psc and estimate the dust mass M$_d$ in the 
usual way (e.g., Zuckerman 2001) from:
\begin{equation}
	M_d = S_{\nu}D^2/ k_{\nu}B_{\nu}(T_d)         
\end{equation}
Here D is the distance between Earth and BP Psc and k$_{\nu}$  is the 
dust opacity.  At submillimeter wavelengths the Planck function B$_{\nu}$ 
can be written as 2kT/$\lambda$$^2$.  Following previous authors, we 
assume that k$_{\nu}$ = 1.7 cm$^2$ g$^{-1}$ at 880 $\mu$m, while recognizing 
that this value may be somewhat on the high side of the true value of 
k$_{\nu}$ (see the motivation for this choice of k$_{\nu}$ by Zuckerman \& 
Becklin 1993 where the value 1.7 cm$^2$ g$^{-1}$ at 800 $\mu$m wavelength was 
first introduced).  This relatively large 880 $\mu$m dust opacity is 
carried by dust particles with radii of a few hundred microns according 
to the study by Pollack et al. (1994).  The absence of a strong silicate 
feature in the 10 $\mu$m window (Fig. 5) is consistent with the 
presence of large particles.  

As noted in Section 4.3.2, BP Psc has little reddening (or bluing). 
Plausibly, optical light seen from BP Psc has pursued a complex path 
through the dusty envelope, undergoing both absorption and scattering.  
If, in addition, there is a wide range of particle sizes, from smaller to 
larger than the wavelengths of interest, then the dominant particles at 
visual wavelengths are apt to have sizes comparable to these wavelengths 
and changes in color should be minimal.

In part because there is no precedent for a giant star with an SED like 
that of BP Psc, we estimate dust masses by considering some results from 
debris disk studies of main sequence stars. The SED of BP Psc indicates 
that orbiting dust is present over a wide range of temperatures.  In 
addition to the hot dust at $\sim$1500 K and warm dust at 210 K, we can be 
quite confident that a substantial mass of cold dust grains is also 
present.  Here we assume the model of a 10 Myr old, 5000 K star, $\sim$100 
pc from Earth.  In this case the semimajor axis of the molecular (CO) disk 
is 60 AU.  Blackbody grains this far from BP Psc that are irradiated by 
the unattenuated stellar radiation field will be at 36 K, while 210 K 
blackbody grains will orbit only $\sim$2 AU from BP Psc.  It is hard to 
conceive of a situation where all the dust would be located so much closer 
to the star than the molecular gas.  Thus, in addition to the 1500 K and 
210 K dust particles, there should also be dust at temperatures too cold 
to have been detected by IRAS.  We can estimate the relative amounts of 
warm and cold dust in the following way.

Williams and collaborators (Williams et al. 2004; Najita \& Williams 
2005; Williams \& Andrews 2006) have measured the 850 $\mu$m flux density 
from 8 stars (HD 8907, 14055, 15115, 21997, 107146, 127821, 206893, 
218396) with IRAS measured far-IR excess emission.  We fitted blackbodies 
to the IRAS fluxes for these 8 stars and compared the expected blackbody 
flux density at 850 $\mu$m to the Williams et al. measurement.  In each 
case, the measured 850 $\mu$m flux density fell below the blackbody flux 
density, by factors ranging between 2.5 and 6, and, on average, by about 
a factor of 4.  From Fig. 6, we see that the 880 $\mu$m flux density of 
BP Psc lies within a factor of two of the plotted 210 K blackbody line.  
The most natural explanation of such an elevated 880 $\mu$m flux density 
is the presence of cold dust not seen by IRAS.  An unambiguous example of 
this phenomenon is Hen3-600 in the TW Hya Association where the 850 
$\mu$m flux density lies well above the Rayleigh-Jeans blackbody 
extrapolation of the far-IR emission (Fig. 2 in Zuckerman 2001).

Based on the preceding discussion, for estimation of dust masses, we will 
assume that 50\% of the 880 $\mu$m flux density (i.e. 9 mJy) is generated 
by dust particles at 210 K and the other 9 mJy by cold dust at 36 K.  
Since some or even all of the 8 Williams et al. stars likely are orbited 
by some cold dust responsible for some of the 850 $\mu$m emission, this 
50-50 division of the 880 $\mu$m flux density is probably somewhat overly 
generous in favor of warm dust at BP Psc.  

With k$_{\nu}$ = 1.7 cm$^2$ g$^{-1}$ and a flux density of 9 mJy at 880 
$\mu$m, for a 10 Myr old star 100 pc from Earth, the mass of dust at 36 K 
is 4.3 $\times$ 10$^{27}$ g or 0.7 Earth masses.  For the 3 Myr and 20 Myr 
old stars in Table 6, this mass would be roughly a factor of two larger 
and smaller, respectively.  At each age, if the flux density carried by 
210 K dust is also 9 mJy, then the mass of 210 K dust is about a factor of 
6 smaller than the mass of 36 K dust.

If instead, BP Psc is a giant star at 300 pc, then the implied disk radius 
would be $\sim$180 AU and dust masses would be an order of magnitude 
larger, or 7 Earth masses.  In addition, as mentioned above, k$_{\nu}$ = 
1.7 cm$^2$ g$^{-1}$ may somewhat overestimate the dust opacity at 880 
$\mu$m, in which case the dust mass would be even larger.  For example, 
for the perhaps somewhat similar giant star HD 233517, Jura (2003) adopted 
an 880 $\mu$m opacity about 5 times smaller than 1.7 cm$^2$ g$^{-1}$.  
(See Section 4.10 for additional discussion of HD 233517.)

\subsection {Gas mass and gas-to-dust ratio}

As noted in Section 3, the upper limit to the $^{13}$CO, J = 
2$-$1 line intensity implies that the 
$^{12}$CO opacity is $<$14 if the isotopic ratios are solar (1/89).  The 
$^{12}$CO optical depth can also be estimated by comparing the peak 
brightness temperature of the J = 2$-$1 and 3$-$2 lines; the latter should be 
about twice as large when both lines are thin and the excitation 
temperatures and source sizes are the same for both transitions.  Another 
potential diagnostic of CO optical depth is comparison of the peak 
brightness temperature of the CO lines with the temperature of the dust 
grains at a given distance from BP Psc; the latter temperature can be 
calculated assuming that the grains are sufficiently large to radiate 
like blackbodies and the grains see the unattenuated heating flux from BP 
Psc.  If the CO lines are optically thick and if the gas is heated only 
by collisions with the dust (but Qi et al. 2006 argue for extra heating 
of the gas relative to the dust at a given distance from TW Hya), then 
the CO brightness temperature and dust temperature should be the same.

In Section 4.6 we noted that the blackbody dust temperature 60 AU from BP 
Psc would be 36 K if these dust particles see the unattenuated heating 
flux from the star.  Notwithstanding the large quantity of warm dust close 
to BP Psc, this assumption could be reasonable given that the disk is 
likely flared.  Specifically, if the total mass of 36 K dust (0.7 Earth 
masses) estimated in Section 4.6 is carried by particles with radii 
$\sim$100 $\mu$m, and if all these grains are located out near 60 AU, then 
they will all see unattenuated starlight if their vertical extent above 
the disk plane subtends only a modest few AU.  While we cannot rule out 
the possibility of some cold dust ``hidden" in the disk midplane closer to 
BP Psc than 60 AU, the feeble 880 $\mu$m flux density guarantees that the 
BP Psc disk carries much less dust mass than disks around most well known 
classical T Tauri stars.  For example, the dust mass at the 8 Myr old TW 
Hya is $\sim$30 times greater than at BP Psc and the dust mass of some Myr 
old classical T Tauri stars is greater by yet another order of magnitude.  
The virtue of measurements in the optically thin 880 $\mu$m continuum is 
they reveal most of the mass in dust particles with radii less than about 
a centimeter.  Somewhat similar arguments would apply if BP Psc is a giant 
star 300 pc from Earth.  For example, Jura (2003) models HD 233517 as a 
flared disk.

The peak CO source brightness temperatures in the 2$-$1 and 3$-$2 lines are 
$\sim$23 K (right hand column of Table 3).  Given uncertainties in the CO 
source sizes (especially for the 2-1 line) and concomitant brightness 
temperature uncertainties, it is not possible to draw any strong 
conclusions about CO optical depths from the various comparisons 
mentioned two paragraphs above other than to say that the $^{12}$CO 
optical depths appear to be not much less than unity.  Future 
interferometric observations of the CO lines with higher spatial 
resolution than those presented here might go a long way in clarifying 
the issue of optical depth.  Detection of $^{13}$CO would also be 
helpful, although this will not be easy given the long integration time 
(490 minutes) devoted to the $^{13}$CO J = 2$-$1 line at the 30\,m telescope.

We estimate a lower limit to the mass in H$_2$ molecules (in solar 
masses) using equations in Scoville et al. (1986).  For CO, their 
expression reduces to: 
\begin{equation}
M_{H_{2}} ~=~ A\times10^{-16} M_{\odot} [T_{x} + 0.92]
(e^{B/T_{x}})
\left( \frac{\tau}{1 - e^{-\tau}} \right) \frac{S_{CO}(Jy ~km s^{-1})D^2(pc)}
{X_{CO}}.
\end{equation}
where the quantities A and B take the values 1.43 and 16.6, respectively, 
for the J = 2$-$1 line and 0.28 and 32.2 for the J = 3$-$2 line.

If BP Psc is a T Tauri star, to derive H$_2$ masses we adopt excitation 
temperature T$_x$ = 36 K, D = 
100 pc, and X(CO) = [CO]/H$_2$ = 10$^{-4}$.  Assuming optically thin 
lines, both the 2$-$1 and 3$-$2 integrated CO line intensities imply H2 
column 
densities $\sim$4 $\times$ 10$^{20}$ cm$^{-2}$ and corresponding H$_2$ masses 
$\sim$10$^{-5}$ M$_{\odot}$ or about 4 Earth masses.  The masses could be an 
order of magnitude larger if the $^{12}$CO J = 2$-$1 line has an optical 
depth $\sim$10 as permitted by the $^{13}$CO J = 2$-$1 non-detection, and 
larger still if the CO is depleted relative to H$_2$ compared to the 
interstellar ratio X(CO) (which we have taken to be 10$^{-4}$).  In any 
event, 4 Earth masses should represent a firm lower limit to the H$_2$ 
mass, excepting the (minor) caveat that X(CO) might be as large as 1/4000 
(Lacy et al. 1994) rather than 1/10000.

Comparison of H$_2$ mass with the cold dust mass of 0.7 Earth masses 
$\sim$60 AU from BP Psc (Section 4.6), yields a gas to dust ratio by mass 
in the range between 6 and about 100.   Within a few AU of BP Psc, where 
the 210 K dust resides, the gas-to-dust ratio is unknown.

If BP Psc is a giant star 300 pc from Earth, then the lower limit to H$_2$ 
mass would be about 40 Earth masses.  Because the dust and gas masses both 
scale as distance from Earth squared, to first order the gas to dust ratio 
by mass would be the same as indicated in the preceding paragraph.  In the 
binary model sketched in Section 4.5, one envisions an 0.1 M$_{\odot}$ star 
spiraling inward by $\sim$5 x 10$^{11}$ cm.  For material contained in a 
concomitant excretion disk of radius 180 AU, conservation of angular 
momentum implies a disk mass of order that of Jupiter, in good agreement 
with the observations provided that the CO J = 2$-$1 line is somewhat 
optically thick.

\subsection {Mass accretion rate}

Mass accretion onto BP Psc is implied by the large quantity of hot dust, 
optical veiling, and strong, broad (full width at 10\% of peak flux 
$\sim$500 km s$^{-1}$) H$\alpha$ emission (Fig. 9).  The mass accretion 
rate is $\sim$10$^{-8}$ solar masses per year based on the H$\alpha$ 10\% 
width and the relationship for pre-main sequence stars shown in Fig. 3 of 
Natta et al (2004).  An independent accretion rate can be estimated 
with the prescription given in White \& Hillenbrand (2004).  As indicated 
in Section 4.3.1, we estimate the ratio of excess continuum flux to 
photospheric flux to be $\sim$0.4 or less in the spectral region near 6500 
\AA .  (This ratio is designated ''r'' by White \& Hillenbrand 2004).  
Then, for a 10 Myr old star of 5000 K, the mass accretion rate in solar 
masses per year is $\sim$10$^{-8}$ and the accretion luminosity is 
$\sim$25\% of the underlying stellar luminosity (White \& Hillenbrand 
2004; R. White personal comm. 2007).  This would be a very large accretion 
rate for a star as old as 10 Myr; with the possible exception of TW 
Hya, we know of no other as large.  Mass loss 
rates for stars ejecting bipolar jets, as per BP Psc, are typically 
estimated to be a few percent of the mass accretion rates (Hartigan et al. 
1995; J. Muzerolle and R. White 2007, personal communications).

Should BP Psc be a post-main sequence giant star, and if the excreted 
disk mass is comparable to the mass of Jupiter (Section 4.7),
then with an accretion rate of 10$^{-8}$ M$_{\odot}$ per year, the disk
might last only 10$^5$ years (see also Section 4.10.2).

\subsection {A Bipolar Herbig Haro Outflow from BP Psc}

Figures 10-12 reveal an extensive chain of Herbig-Haro (HH) objects, the 
shock-excited emission nebulae that trace mass outflows from young stars. 
The bipolar outflow extends to the northeast (position angle 
$\sim$24$^{\circ}$) and southwest of BP Psc, consistent with the 
orientation 
of the polar axis of the circumstellar disk seen in our Keck AO 
observations.  As discussed in Section 3, we deduce that the side of the 
disk nearest to Earth is to the SW and, thus, we conjecture that the 
northeast 
jet is probably the redshifted side, while the southwest jet is blueshifted.

The overall appearance of the outflow is roughly symmetrical; on each 
side of the central star, a narrow linear filament stretches several 
arcminutes away from the star, beyond which one or more bow shocks are 
visible. The filaments are clumped and knotty, like beads on a string, 
consistent with internal shocks due to turbulence or velocity variations 
within the outflow.  Table 7 and Figure 11 give the positions of the 
various components of the HH 999 outflow.  But the symmetry breaks down 
upon closer inspection, revealing extensive differences between the two 
sides in shape and brightness. The northern HH objects are brighter and 
more extensive than the southern ones, and are visible to a greater 
distance from the star.  The bright southern bow shock, knot S4, is 
$\sim$2.7$\arcmin$ from BP Psc, while the prominent northern shocks N3-N6 are 
$\sim$3.25\arcmin, 3.95\arcmin, 4.25$\arcmin$ and 6.50$\arcmin$ from the star. 
There are faint 
suggestions of additional nebulosity at greater distances from the star, 
(knots S5, S6, and N7 in Table 7), but deeper integration is necessary to 
confirm these low S/N detections.

At an estimated distance of $\sim$100 pc, the observed jet extends at 
least 0.25 pc, or 0.5 pc if the faint outer HH objects are confirmed.  
Its total extent may well be greater than this, extending outside our 
current field of view, since many T Tauri stars are known to launch 
multi-parsec-long outflows (McGroaty \& Ray 2004, and references 
therein). Deep observations of the region around BP Psc with larger 
fields of view will be required to determine the true physical size of 
the outflow.

The two sides also differ in the relative intensities of the H$\alpha$ 
and [S II] emission lines (Figs. 10 \& 11): the southern filament S3 is 
much brighter in [S II] than H$\alpha$, while the northern filament N1 is 
comparably bright in both. The two knots immediately southwest of BP Psc, 
S1 and S2, are visible in [S II] emission only. With the exception of 
these knots and the southern filament, the H$\alpha$ emission generally 
extends 
over larger angular scales than the [S II] emission. In particular, for 
all the bow shocks, the leading edges are visible in H$\alpha$, while the 
[S II] is confined to further back behind the shock. This is entirely 
consistent with the bow shock plus Mach disk emission features of other 
similar outflows such as HH 34 (Reipurth et al. 2002).  

Once away from the star, the NE and SW sides of the jet are not separated 
by 180$^\circ$, because the outflow overall is curved, not straight (Fig. 
11, third panel). Close to the star, the inner filaments appear well 
collimated, but at larger distances the HH objects appear increasingly 
deflected toward the northwest. The three outer, low S/N HH objects, 
knots S6, S7 and N8 (outside the field of view of Fig. 11) continue 
this curving pattern, with position angles increasingly far from the 
initial jet axis (Table 7). While precession has been invoked to explain 
curvature in other HH jets, that mechanism generally produces 
symmetrically curved, S-shaped outflows, in contrast to the morphology 
observed here. Instead, the BP Psc flow appears to be part of the rarer 
class of ``C-shaped'' outflows such as HH 334 and 366 (Bally et al. 1996; 
Bally \& Reipurth 2001). It has been suggested that such shapes arise 
because of deflection of jets due to proper motion between the star and 
the surrounding nebula (i.e. a crosswind; see Lebedev et al. 2004).  The 
proper motion of BP Psc (Section 4.2) toward the SE is in a direction 
consistent with this model.   From Figure 11, one deduces that, in the 
plane of the sky, the jet velocity is $\sim$17 times the velocity of the 
star, if the jet material were to completely lose the proper motion of 
the star at the outset of the outflow.  This gives an upper limit to the 
jet velocity of $\sim$390 km s$^{-1}$ (at an assumed distance of 100 
pc).  Since, in the model, the loss of the component of motion toward the 
SE is due to the crosswind, the actual jet velocity is certainly less 
than 390 km s$^{-1}$ and can be derived from measurements of proper 
motions of the HH objects along the direction of the jet.

After the discovery of the jet, we re-examined the red plates from the 
Digitized Sky Survey\footnote{The Digitized Sky Surveys were produced at 
the Space Telescope Science Institute under U.S. Government grant NAG 
W-2166. The images of these surveys are based on photographic data 
obtained using the Oschin Schmidt Telescope on Palomar Mountain and the 
UK Schmidt Telescope.}, and discovered that several components of the BP 
Psc jet complex are visible in those data (Fig. 11, bottom panel). This 
presented an 18-year temporal baseline to check for proper motion of the 
clumps. Knot S4 (which is the brightest portion of the nebula, and thus 
easiest to detect in the DSS) does appear to move between the two epochs. 
Its apparent shift of $\sim$5.1$\arcsec$ over 18 years implies a proper 
motion of 0.28$\arcsec$ yr$^{-1}$, corresponding to 125 km s$^{-1}$ at a 
distance of 100 pc or 375 km s$^{-1}$ at 300 pc.  We caution, however, 
that the proper motion is a very tentative 
measurement, given the uncertainty in the distance to BP Psc, the limited 
signal-to-noise on the clumps in the DSS image, and the possibility of 
systematic offsets between the DSS and Lick PFCam astrometric solutions.

High velocity jets are of course known from many pre-main sequence stars, 
but also from a modest number of evolved, high-luminosity, protoplanetary 
and planetary nebulae (e.g. Sahai 2002).  One of the more striking examples 
of the evolved class is Henize 2-90 (Sahai 2002, Sahai et al. 2002, Kraus et 
al 2005, Garcia-Segura et al 2005) that has been analyzed in various ways in 
various papers involving some combination of general mass outflow, binarity, 
an accretion disk, magnetic fields and rapid rotation.  He2-90 is of much 
earlier spectral type than BP Psc and the observed linear extent of the 
He2-90 outflow (even taking into account the uncertain distances to both 
objects) is much less than that at BP Psc.

Due to the visual obscuration near BP Psc, the launch region for its jets 
is better studied at radio wavelengths.  Reipurth \& Bally (2001) 
summarize interferometric observations of centimeter wavelength continuum 
emission at young stars with associated jets.  Accordingly, we used the 
VLA (see Section 2.1) in an imaging search for X (3.5 cm) and L (20 cm) 
band emission, in part to learn more about the launch region of the 
jets.  An additional motivation for these observations was measurement of 
the parallax of BP Psc.   Loinard et al. (2005) demonstrate that, for T 
Tauri stars accompanied by moderately strong radio sources as per T Tauri 
itself, the VLBA can be employed to measure stellar distances with 
exquisite precision.  Unfortunately, we measured only upper limits to the 
BP Psc flux densities: $<$120 $\mu$Jy/beam at X-band and $<$1.7 mJy/beam 
at L-band.   Hence, parallax measurements will require optical images.

\subsection {Origin of BP Psc}

\subsubsection {T Tauri star}

A characteristic of most known classical T Tauri stars is 
association with interstellar nebulosity.  In such cases, where stars are 
only a few Myr old, we know the stellar birthplace.  However, by $\sim$10 
Myr, interstellar nebulosity has generally dissipated and usually other 
methods must be used to deduce the birthplace of rare, surviving, 
classical T Tauri stars and the more common, weak-lined, post T Tauri 
stars.  Thus, for example, TW Hya, an isolated 8 Myr old classical T 
Tauri star, likely had its origin in the Lower Centaurus-Crux region 
(e.g., Section 6 in Zuckerman \& Song 2004).

In contrast, if BP Psc is a pre-main sequence star, then its birthplace is
unclear.  Unlike TW Hya, which belongs to an association with dozens of
known members, so far we have been unable to identify any stars in the
vicinity of BP Psc with similar Galactic space motions and ages $\leq$10
Myr.  The center of the nearest known prominent interstellar cloud MBM 55
(e.g., Hearty et al. 1999) is $\sim$20$^\circ$ north of BP Psc, and MBM 55
is devoid of any evidence of recent star formation (L. Magnani, private
comm. 2007).  In addition, tracing the proper motion of BP Psc back for 10
Myr brings it no closer to MBM 55 than it is now (with the caveat that the
proper motion of MBM 55 is unknown due to lack of associated stars).

\subsubsection {Post-main sequence star}

If BP Psc is a post-main sequence star, then one plausible model is that
BP Psc was once a close binary star, perhaps of the W UMa class, and the
secondary was consumed by the primary after the primary left
the main sequence.  Alternatively, the secondary might still be 
present (Section 4.5), and responsible for ejection of material from a 
common envelope formed with BP Psc.  Rucinski (2006) estimates that W UMa 
stars comprise
perhaps one in 500 of main sequence FGK stars.  The lifetime of a 1.8
solar mass star while a first ascent red giant is a few times 10$^8$ years
(Iben 1991).  If the disk mass is $\sim$0.001 M$_{\odot}$ (Section 4.7), 
at
the current accretion rate of 10$^{-8}$ M$_{\odot}$ per year, the gaseous
disk might last only 10$^5$ years.  Then we might expect to see the BP
Psc phenomenon in about one first ascent giant star in one million.

In a search of more than 40,000 luminosity class III giants, Zuckerman et
al (1995b) found none anywhere near as infrared bright as BP Psc.  Melis
et al (in preparation) searched $\sim$100,000 giant stars in the Tycho
catalog for excess infrared emission and found BP Psc (if it is a giant)
and another star of spectral type F$-$with some (but not all)
characteristics similar to those of BP Psc$-$that may also be a first
ascent giant star.  Earlier, Jura (2003) modeled the dusty first ascent
red giant HD 233517 as arising from engulfment of a low-mass companion
star.  However, at HD 233517, there is no direct evidence of a gaseous
disk, hot dust, mass accretion, or mass outflow.  Also, the dust around HD
233517 is rich in carbon-rich PAH components (Jura et al 2006), for which
there is no evidence in the 10 $\mu$m spectrum of BP Psc (Fig. 5).  
Finally, the CO images of BP Psc indicate a gaseous disk radius of
$\sim$180 AU if the star is 300 pc from Earth, while the Jura model for HD
233517 suggests that the radius of its dusty disk may be $\leq$50 AU.

Soker (1998) presents arguments in support of the binary progenitor model
(e.g., Morris 1987) for bipolar planetary nebulae.  Soker's expectation is
that a large fraction of the first ascent giant branch descendants of
low-mass ($M \stackrel{<}{\sim} 2$ $M_\odot$) stars in close (separation
$\stackrel{<}{\sim} 10$ AU) binary systems will interact with their
companions and experience a common envelope phase that is accompanied by
intense mass-loss.  Although this model does not directly predict that
such a high-mass-loss-rate first ascent giant phase would involve the
formation of a disk and jets, such structures nevertheless might be
expected, given the close analogy with binary interactions (and resultant
axisymmetric mass loss) following the AGB (e.g., Nordhaus \& Blackman 
2006).

\section {Conclusions}

We have studied the optically variable star BP Psc with a variety of 
observational techniques spanning radio to optical wavelengths.  This 
previously neglected star might be one of the nearest and oldest known 
classical T Tauri stars.  However, more likely, it is the first known 
example of a first ascent, post-main sequence giant star with an 
associated orbiting massive molecular disk, rapid gas accretion, and 
outflowing bipolar jets and Herbig-Haro objects.  The disk may be the 
aftermath of accretion of an erstwhile or extant low mass companion of BP 
Psc, with material excreted from a common envelope as the companion is 
enveloped by BP Psc as the latter expands during the first ascent giant 
phase.  Whatever the disk origin, planets may now be forming in it, a Gyr 
or more after the formation of BP Psc itself.

Additional measurements are obviously desirable including
frequent monitoring of radial velocities to establish whether or not BP 
Psc is a close binary, X-ray flux and spectrum, extensive study of the 
bipolar jets, high-resolution aperture synthesis of CO emission, and high 
resolution AO and/or HST imaging in the near-IR.  Most important, a 
measurement of trigonometric parallax should clarify the evolutionary 
state of BP Psc.  However, because of uncertainty associated with mass 
accretion luminosity and preferential viewing orientation nearly in the 
disk plane, even when we know how far BP Psc is from Earth, we will not 
know exactly how luminous the underlying star is.

\acknowledgements 

We thank Emily Rice, Greg Wirth, and Ian McLean for helping to obtain the 
epoch 2006 NIRSPEC data and Ms. Rice for aid in their reduction.  We 
acknowledge the efforts of Richard Webb in obtaining epoch 1996 data and 
thank Frank Fekel and Russel White for contributions.  We are very grateful 
to Katia Biazzo and Antonio Frasca for important help in classifying BP Psc.  
We thank the referee for constructive comments, especially for a suggestion 
to consider gravity-dependent line ratios.  This research was supported in 
part by NASA grants to UCLA.

\clearpage

\begin{deluxetable}{llllll}
\tablenum{1}
\tablewidth{0pt}
\tablecolumns{6}
\tablecaption{CO Observations with the JCMT and IRAM 30\,m Telescopes}
\tablehead{

\colhead{Telescope} 
&\colhead{Date} 
&\colhead{FWHP}
&\colhead{Molecule} 
& \colhead{Transition} 
& \colhead{Frequency} \\ 
\colhead{} 
&\colhead{(UT)} 
&\colhead{(arcsec)}
&\colhead{}
&\colhead{} 
& \colhead{(MHz)} 
}

\startdata
JCMT & Feb 1996 & 14 & CO & J = 3-2 & 345796.0 \\
IRAM & May 1996 & 11 & CO & J = 2-1 & 230538.0 \\
  (30 m) & June 1996 & 11 & $^{13}$CO & J = 2-1 & 220398.7 \\  
\enddata

\end{deluxetable}

\clearpage

\begin{deluxetable}{lcccccc}
\rotate
\tablenum{2}
\tablewidth{0pt}
\tablecaption{OVRO and SMA: System Characteristics}
\tablehead{
\colhead{Map} 
&\colhead{Freq.} 
&\colhead{$\Delta V_{chan}$}
&\colhead{$\Delta \nu_{band}$} 
& \colhead{Beamsize} 
& \colhead{Noise level} 
& \colhead{Detected Flux} \\
\colhead{} 
&\colhead{\it (GHz)} 
&\colhead{($km~s^{-1}$)}
&\colhead{\it (MHz)}
& \colhead{\it (arcsec; deg)} 
& \colhead{\it (K/Jy Bm$^{-1}$)}
&\colhead{(\%)} 
}
\startdata
CO(2-1)\tablenotemark{a}& 230.538 & 0.70 & 56 &$2.35\times 2.23;-75^{o}$ 
& 
0.53/0.12 & 100 \\
CO(3-2) [NA]\tablenotemark{b}& 345.796 & 0.71 & 102.4 
&$2.39\times 1.98;29^{o}$& 0.37/0.17 & 96  \\
CO(3-2) [UN]\tablenotemark{b}&  & & 
&$2.09\times 1.93;63^{o}$& 0.53/0.21 & 90  \\
1.4 mm\tablenotemark{a}& 230.55 & \nodata  & 1000 &$2.30\times 
2.18;-77^{o}$ & 
0.022 /0.0047 & \nodata \\
880 $\mu$m\tablenotemark{b} & 340.886 &\nodata  &4000  
&$2.46\times 1.96;27^{o}$ & 0.048/0.0022 & \nodata \\
\enddata
\tablenotetext{a}{For observations made from 09 March 1996 - 10
January 1997, with a phase center of $\alpha = 23^{h} 22^{m}
24^{s}.739 ~~\delta = -02^{o} 13' 41.^{''}186$ (J2000); V$_{LSR}$=-17
$km ~s^{-1}$ }
\tablenotetext{b}{For observations made from 25 September 2006, with a
phase center of $\alpha = 23^{h} 22^{m} 24^{s}.69 ~~\delta = -02^{o}
13' 41.^{''}399$ (J2000); V$_{LSR}$=$-15$ $km ~s^{-1}$ }
\end{deluxetable}

\clearpage
\thispagestyle{empty}

\begin{deluxetable}{lccccccc}
\rotate
\tablenum{3}
\tablewidth{0pt}
\tablecaption{Observational Data from OVRO and the SMA}
\tablehead{
\colhead{Map}
&\colhead{I$_{pk}$\tablenotemark{a}}
& \colhead{S$_{tot}/$I$_{tot}$\tablenotemark{b}}
&\colhead{$V_{o}$\tablenotemark{b}}
&\colhead{$\Delta V_{1/2}$\tablenotemark{b}}
& \colhead{$\alpha_{o}$(J2000)\tablenotemark{c}}
& \colhead{$\theta_{a}\times\theta_{b}; pa$\tablenotemark{c}}
& \colhead{$^{source}T_{pk}$\tablenotemark{e}} \\
\colhead{}
&\colhead{\it (Jy)}
& \colhead{({\it Jy km s$^{-1}$})}
&\colhead{($km~s^{-1}$)}
&\colhead{\it ($km~s^{-1}$)}
& \colhead{$\delta_{o}$(J2000)}
& \colhead{\it (arcsec; deg)}
& \colhead{\it (K)} \\
\colhead{}
&\colhead{\it (K)}
& \colhead{({\it K $km~s^{-1}$})}
&\colhead{}
&\colhead{}
& \colhead{}
& \colhead{}
& \colhead{}
}
\startdata
CO(2-1) & $1.4\pm0.12$ & $13\pm0.4$ & $-14.9\pm0.2$\tablenotemark{d}
& $7.0\pm0.6$ & 23:22:24.64$\pm$0.02 & $~1.9~\times ~0.73; ~91$ & 24 \\
& $6.2\pm0.5$ & $57\pm2$ & $-20.9\pm0.1$\tablenotemark{d}
& $2.6\pm0.3$ & $-02:13:41.52\pm0.2$ &$\pm0.2~\pm0.2~\pm4$ & \\
1.3 mm & $<$0.0094 &\nodata &\nodata & \nodata& \nodata& \nodata & 
\nodata\\
& $<$0.043 &\nodata &\nodata & \nodata& \nodata& \nodata  \\
CO(3-2) & $3.7\pm0.17$ & $48\pm0.2$ & $-17.6\pm0.13$\tablenotemark{d}
& $10.5\pm0.35$ & 23:22:24.72$\pm$0.01 & $~1.8~\times ~0.80; ~103$ & 22 \\
& $8.0\pm0.37$ & $100\pm0.4$ & & &$-02:13:41.73\pm0.1$
& $\pm0.1~\pm0.1~\pm10$ & \\
880$\mu$m & $0.018\pm0.002$ &\nodata &\nodata & \nodata
& 23:22:24.72$\pm$0.02& 0.81 $\times <0.2; \sim 16$ & \nodata \\
& $0.039\pm0.0044$& & & & $-02:13:41.55\pm0.2$
&$\pm0.2~~~~~~~~~~~~~~~~~~$ & \\
\enddata
\tablenotetext{a}{Uncertainties are those measured from the map noise.
The uncertainties do not include (the larger) absolute flux
calibration uncertainties of $\sim 20\%$. Upper limits are 2$\sigma$.}
\tablenotetext{b}{Uncertainties reflect the statistical uncertainties
of the gaussian fit only.}
\tablenotetext{c}{For the CO(3-2) data fits are based on the uniformly
weighted dataset.  Uncertainties are estimate as $\theta_{bm}\times
(1/SNR)$ and do not include any systematic uncertainties.}
\tablenotetext{d}{The CO(2-1) line was fitted with two gaussian
components.  The two values of each parameter for CO(2-1) represent
each fitted component. All velocities are converted to heliocentric
assuming $v_{Helio}~=~v_{LSR}~-~2.72 ~km~s^{-1}$ for BP Psc.}
\tablenotetext{e}{The implied source brightness temperature based on
the fitted source size.}
\end{deluxetable}

\clearpage

\begin{deluxetable}{lccc}
\tablecaption{BP Psc Michelle Mid-Infrared Photometry}
\tablenum{4}
\tablewidth{0pt}
\tablecolumns{4}
\tablehead{
\colhead{Filter} &\colhead{Central Wavelegth} 
&\colhead{Flux Density} &\colhead{Uncertainty} \\
\colhead{} &\colhead{($\mu$m)} 
&\colhead{(Jy)} &\colhead{(Jy)} 
}
\startdata
Si-1            &7.7       &1.40                &0.10\\
Si-2            &8.8       &1.34                &0.04\\
N$^\prime$      &11.2      &1.89                &0.06\\
Si-5            &11.6      &1.84                &0.06\\
Si-6            &12.5      &2.11                &0.06\\
Qa              &18.1      &5.2                 &0.3\\
\enddata
\end{deluxetable}

\clearpage

\begin{deluxetable}{llcclc}
\rotate
\tablenum{5}
\tabletypesize{\small}
\tablewidth{0pt}
\tablecaption{Journal of HIRES and NIRSPEC Observations}
\tablecolumns{6}
\tablehead{

\colhead{Instrument} 
&\colhead{UT Date} 
&\colhead{Spectral Range}
&\colhead{Spectral} 
& \colhead{H$\alpha$ EW} 
& \colhead{Heliocentric Radial} \\ 
\colhead{} 
&\colhead{} 
&\colhead{}
&\colhead{Resolution}
&\colhead{(\AA)} 
& \colhead{Velocity (km s$^{-1}$)} 
}

\startdata
HIRES & 12 Jul 1996 & 3840-6260 \AA\ & 65816 &   & $-14.74 \pm 1.71$ \\
  & 10 Oct 1996 & 3810-6235 \AA\ & 65296 &   & $-11.46 \pm 1.73$ \\
  & 01 Sep 2006 & 5690-8550 \AA\ & 40646 & $-12.95 \pm 0.30$ & $-15.53 \pm 0.79$ \\
  & 09 Sep 2006 & 5690-8670 \AA\ & 41263 & $-15.20 \pm 0.36$ & $-14.14 \pm 0.83$ \\
  & 15 Oct 2006 & 3720-7992 \AA\ & 51344 & $-11.37 \pm 0.40$ & $-13.20 \pm 1.47$ \\
NIRSPEC & 09 Aug 2006 & 1.50-1.78 $\mu$m & 2000 &   & \\
   &  & 2.01-2.44 $\mu$m & 2000 &  & \\
   &  & 2.83-3.66 $\mu$m & 2000 &  & \\
   &  & 1.15-1.32 $\mu$m & 18750 &   & \\
\enddata

\tablecomments{The 0.79 km s$^{-1}$ radial 
velocity uncertainty given for 1 Sept. 2006 includes all uncertainties 
including those in the 
velocity of the standards (see Section 4.5).  The uncertainties listed 
for 9 Sept. and 15 Oct. 2006 are based on cross correlation with the 1 
Sept. 2006 spectrum and are derived by adding, in quadrature, the 0.79 km 
s$^{-1}$  1 Sept. uncertainty and the uncertainty derived from the cross 
correlation.   Relative to the radial velocity on 1 Sept. 2006, the 
velocity on 9 Sept. was less negative by 1.39 $\pm$ 0.23 km s$^{-1}$ , 
while the difference between the 15 Oct. and 1 Sept. velocity was 2.33 
$\pm$ 1.23 km s$^{-1}$.  Derivation of the 1996 epoch velocities and 
errors is discussed in Section 4.5}
\end{deluxetable}

\clearpage

\begin{deluxetable}{cccccccc}
\rotate
\tablenum{6}
\tablewidth{0pt}
\tablecaption{BP Psc: Pre-main Sequence Evolutionary and Kinematic 
Properties}
\tablecolumns{8}
\tablehead{

\colhead{Age} 
& \colhead{M} 
& \colhead{R}
& \colhead{D} 
& \colhead{U} 
& \colhead{V} 
& \colhead{W}  
& \colhead{D$'$}  \\ 
\colhead{(Myr)} 
& \colhead{(solar masses)} 
& \colhead{(solar radii)}
& \colhead{(pc)}
& \colhead{(km s$^{-1}$)} 
& \colhead{(km s$^{-1}$)} 
& \colhead{(km s$^{-1}$)}
& \colhead{(pc)}}

\startdata
3  & 1.72 & 2.13 & 157 & -21 & -36 & -4 & 121 \\
5  & 1.62 & 1.88 & 139 & -19 & -33 & -2 & 107 \\
10 & 1.26 & 1.41 & 104 & -14 & -26 & +2 & 80  \\
20 & 1.04 & 1.10 &  81 & -12 & -23 & +5 & 62  \\
\enddata
\tablecomments{M and R are based on evolutionary models of Baraffe et al 
(2003) and Baraffe 2006 (personal comm.).   Galactic UVW velocities are 
with respect to the Sun and U is defined to be positive toward the 
Galactic center.   At all ages the stellar temperature is assumed to be 
5000 K.  D is the appropriate distance to BP Psc if the integrated flux 
in Figure 6 is radiated isotropically and none of this flux is due to 
accretion of disk material onto BP Psc.  D$'$ is the estimated distance 
to BP Psc when mass accretion luminosity and anisotropic emission are 
taken into account.  See Section 4.3.1 for details.} \end{deluxetable}

\clearpage

\begin{deluxetable}{cllccl}
\tablenum{7}
\tablewidth{0pt}
\tablecaption{Herbig-Haro Objects}
\tablecolumns{6}
\tablehead{
Name & R.A. & Dec. & Separation & Pos. Angle & Comment \\
     &      &      & (arcsec)   & (degrees)  &          
}
\startdata
N7 & 23:22:38.7 & $-02:03:46$ &  631.3 &   19.4  & very faint      \\
N6 & 23:22:34.5 & $-02:07:38$ &  392.0 &   22.0  & bow shock       \\
N5 & 23:22:31.3 & $-02:09:49$ &  252.6 &   23.1  & bow shock       \\
N4 & 23:22:31.5 & $-02:10:06$ &  238.3 &   25.3  & bow shock       \\
N3 & 23:22:29.9 & $-02:10:43$ &  194.7 &   23.6  &         \\
N2 & 23:22:29.5 & $-02:11:01$ &  175.9 &   24.2  &         \\
N1 & 23:22:28.9 & $-02:11:14$ &  160.3 &   23.2  & filament end    \\
    ''    & 23:22:27.6 & $-02:12:08$ &  103.1 &   25.0  & filament start  \\
    \\
S1 & 23:22:24.4 & $-02:13:48$ &    7.9 & -146.4  &         \\
S2 & 23:22:24.1 & $-02:13:57$ &   17.9 & -150.5  &         \\
S3 & 23:22:22.7 & $-02:14:46$ &   71.1 & -155.2  & filament start  \\
    ''    & 23:22:20.4 & $-02:15:35$ &  130.5 & -150.5  & filament end    \\
S4 & 23:22:19.1 & $-02:16:03$ &  164.5 & -149.4  & bow shock       \\
S5 & 23:22:14.9 & $-02:17:21$ &  264.1 & -146.3  & very faint      \\
S6 & 23:22:10.9 & $-02:18:35$ &  359.1 & -144.9  & very faint      \\
\enddata
\tablecomments{Herbig Haro objects observed around BP Psc, ordered 
from north to south. For the long narrow filaments, the coordinates of
each end are given. The curvature of the jet is apparent in how the 
position angles deflect with increasing distance from BP Psc. The
outermost knots in both the north and south directions are detected at 
low S/N in our images and would benefit from additional
observations.}
\end{deluxetable}

\clearpage

\begin{figure}
\plotone{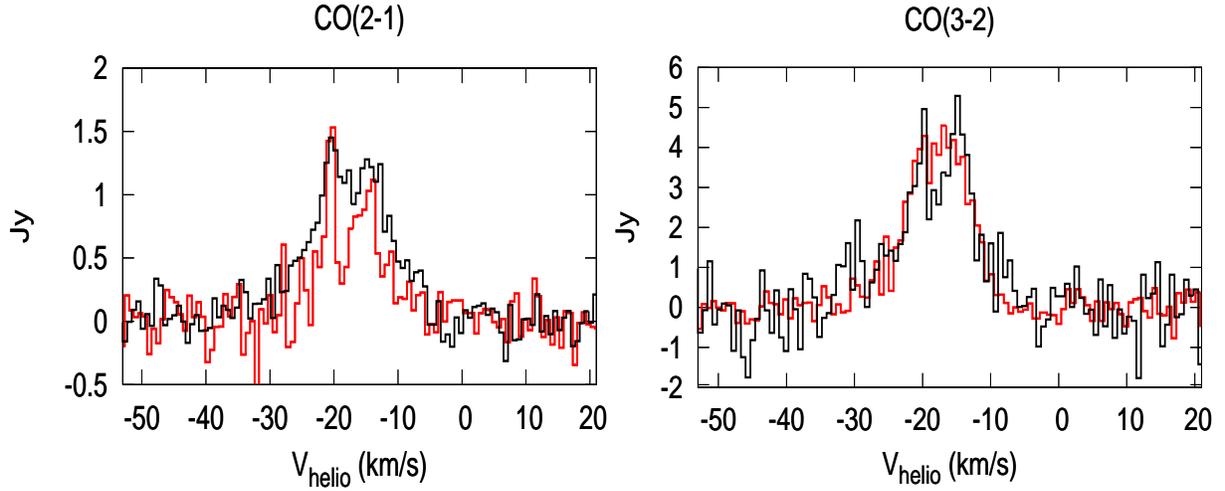}
\caption{Single dish and interferometer CO spectra of BP Psc.   
{\it Left panel}: Spectra for CO(2-1) from IRAM 30 m (black line) and 
OVRO (red line).  {\it Right panel}:  CO(3-2) with the JCMT (black) 
and SMA  (red).  All spectra are put on the same flux scale 
(assuming 3.9 Jy/K for IRAM and 15.6 Jy/K for the JCMT) and 
velocity scale V$_{helio}$ = V$_{lsr} - $2.72 km s$^{-1}$.}

\end{figure}

\clearpage

\begin{figure}
\plotone{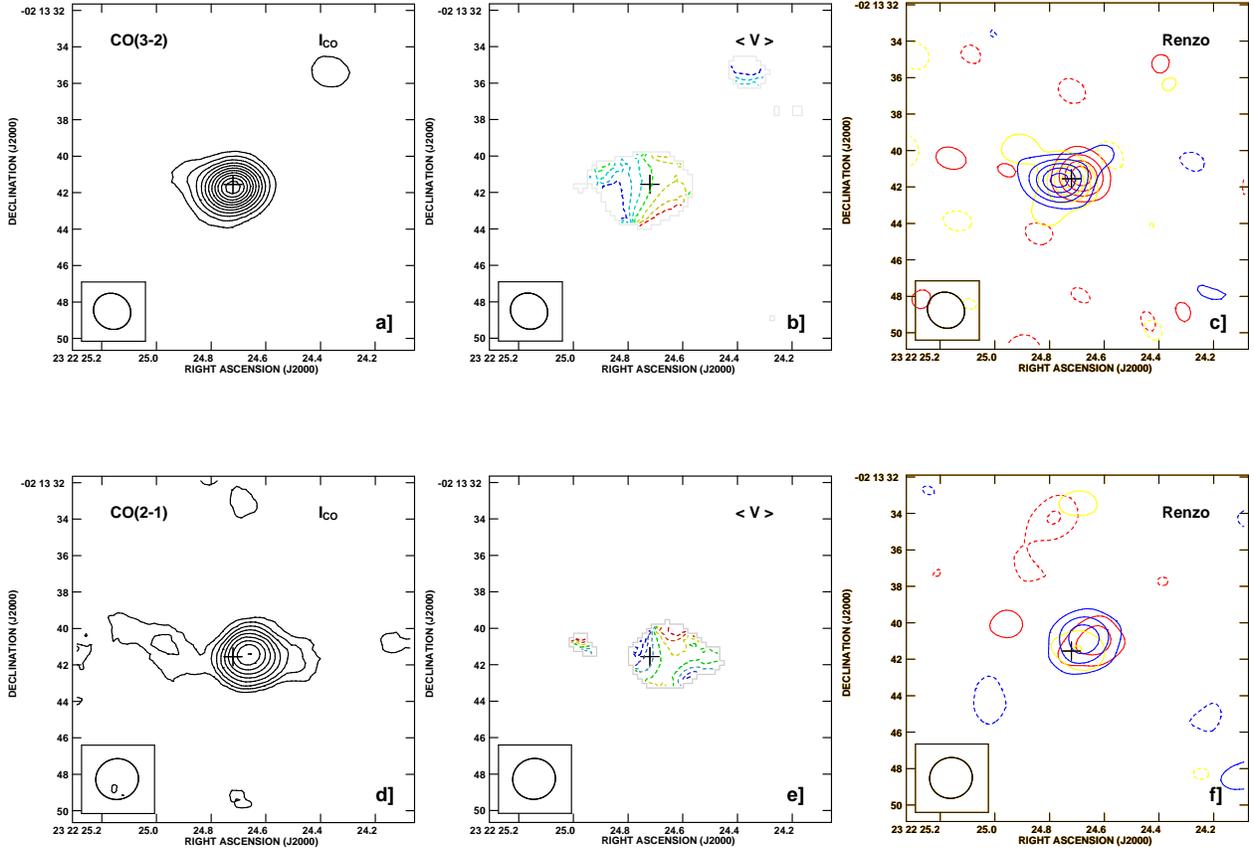}
\caption{{\it a)} The CO(3$-$2) uniformly weighted integrated 
intensity map.  Contours are in steps of 2.75 Jy bm$^{-1}$ 
km s$^{-1}$ (7.0 K km s$^{-1}$).  {\it b)} The CO(3$-$2) velocity 
centroid map.  Contours are in steps of 0.4 km s$^{-1}$ ranging 
from $-18.72$ km s$^{-1}$ (blue) to $-16.32$ km s$^{-1}$ (red) 
(heliocentric).  {\it c)} The CO (3$-$2) renzogram.  Red contours 
are the $-14.7$ km s$^{-1}$ channel, yellow the $-17.6$ km s$^{-1}$ 
channel, and the blue $-20.4$ km s$^{-1}$ (heliocentric).  Contours 
increments are three times the noise levels (sigma) listed in 
Table 2. {\it d)} As in ({\it a}) except for CO (2$-$1).  Contours are 
in steps of 1.0 Jy bm$^{-1}$ km s$^{-1}$ (4.4 K km s$^{-1}$).  
{\it e)} As in ({\it b}) except for CO (2$-$1).  Contours are in steps 
of 0.33 km s$^{-1}$ starting at -18.66 km s$^{-1}$ (heliocentric).   
{\it f)}  As in ({\it c}) except for CO (2$-$1).  Red contours are the 
$-14.6$ km s$^{-1}$ channel, yellow the $-17.3$ km s$^{-1}$ channel, 
and the blue $-20.1$ km s$^{-1}$ (heliocentric).  Contours intervals 
are three sigma.  The cross in each frame marks the fitted position 
of the 880 $\mu$m continuum source.  The beam is displayed at the 
lower left of each frame.}
\end{figure}

\clearpage

\begin{figure}
\plotone{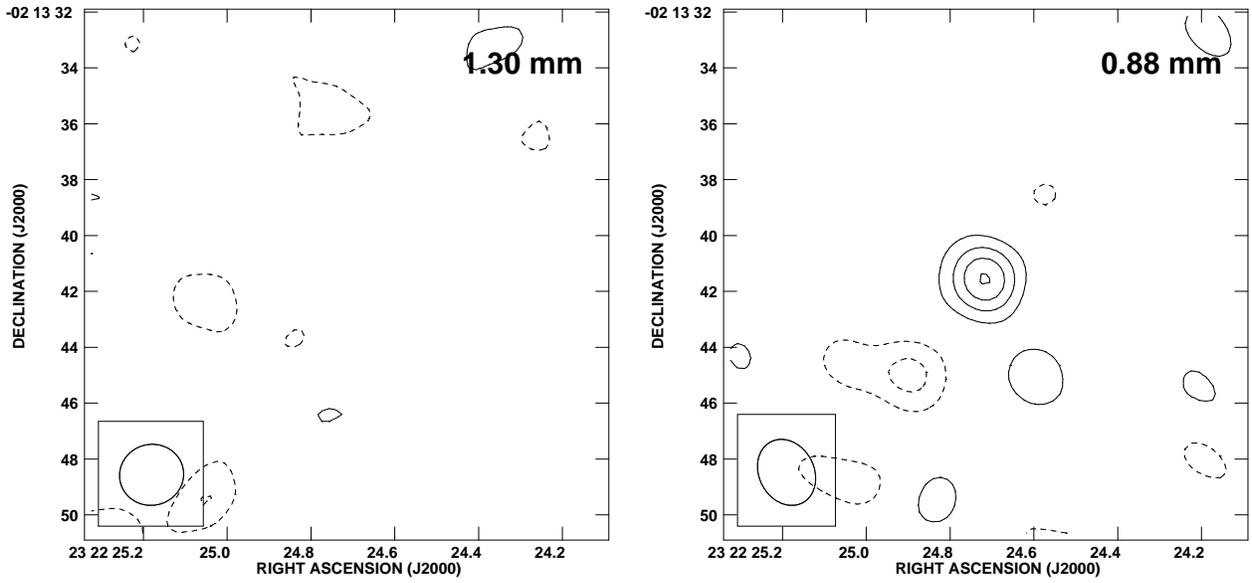}
\caption{Radio continuum images of BP Psc.  Left panel:  The OVRO 
230.6 GHz image with contours in steps of 9.4 mJy bm$^{-1}$ or 
two sigma.  The beamsize is plotted in the lower left corner.  
Right panel:  The SMA 340.9 GHz continuum image with contours 
in steps of 4.4 mJy bm$^{-1}$ or two sigma.}
\end{figure}

\clearpage

\begin{figure}
\epsscale{1.0}
\plotone{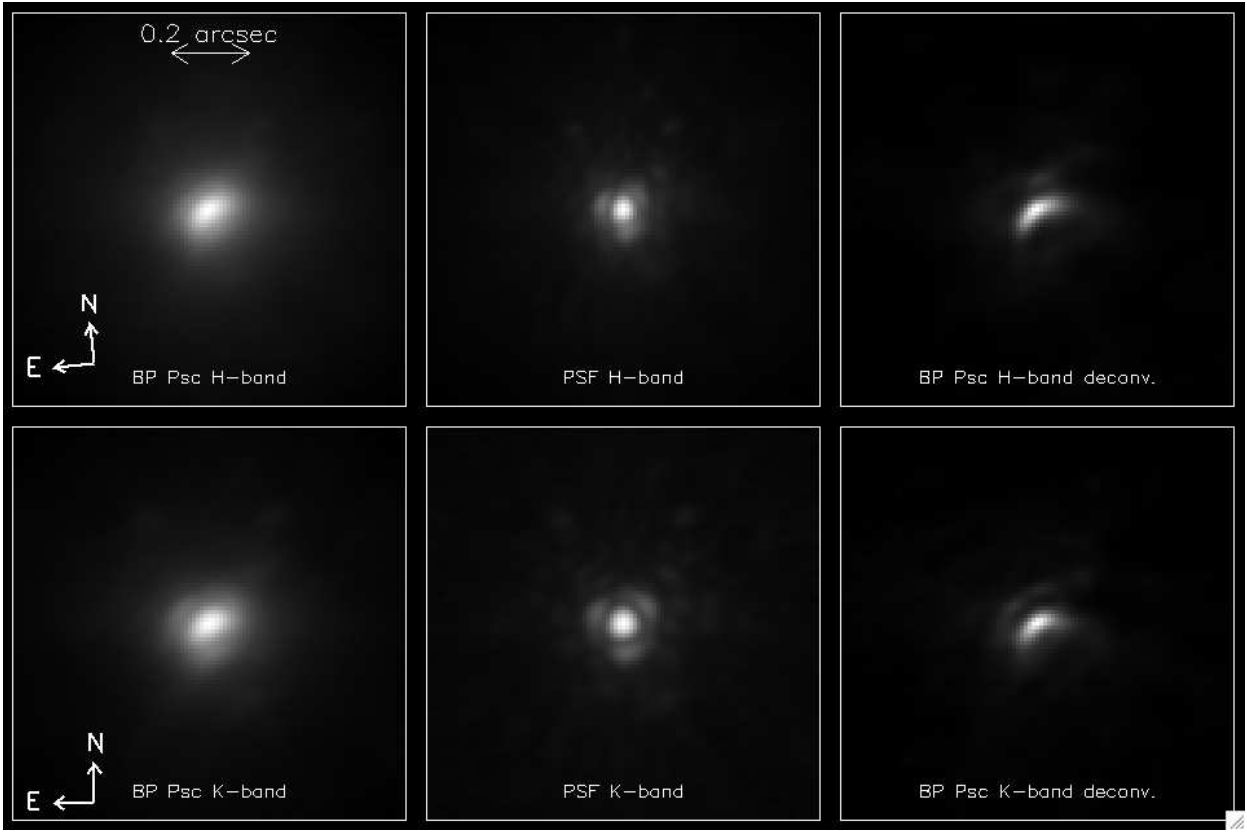}
\caption{H- and K$'$-band AO 
images of BP Psc and a nearby PSF reference star  (TYC 5244-226-1).  
The top row H-band images, left to right, are BP Psc, TYC 5244-226-1, 
and the image of BP Psc deconvolved with a Lucy-Richardson algorithm.   
Similarly for the bottom row, except at K$'$. All images are one arc 
second on a side and are presented on a square-root intensity 
stretch.   Pixel size is 0.01 arcseconds.}
\end{figure}

\clearpage

\begin{figure}
\plotone{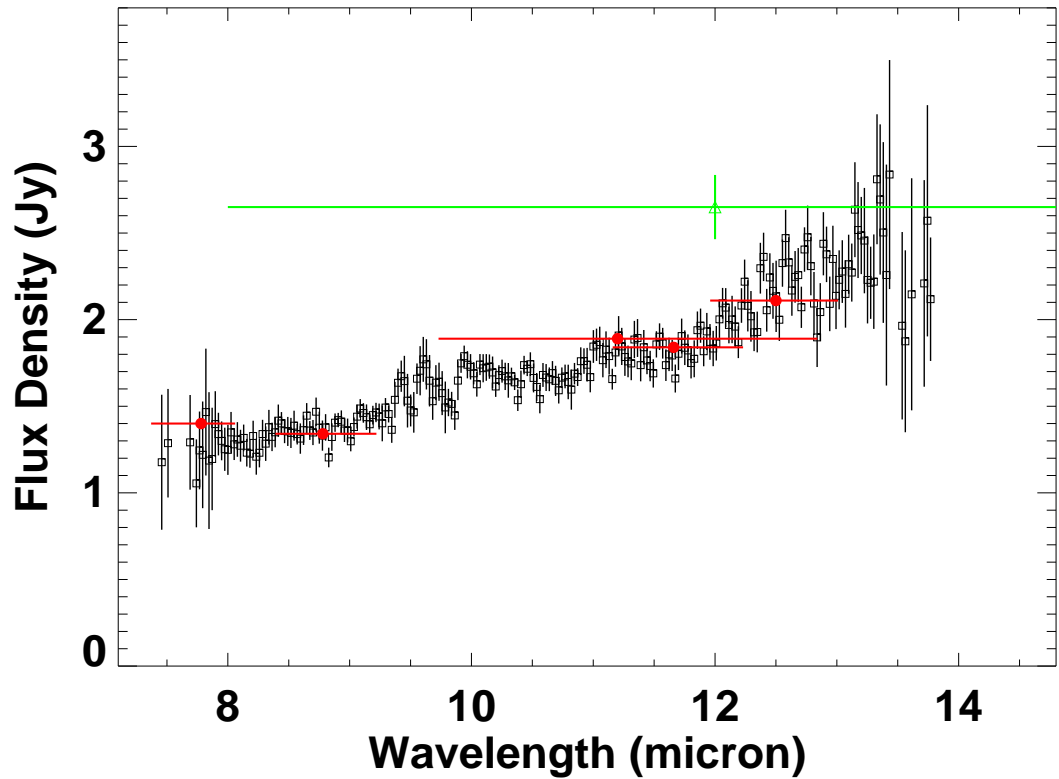}
\caption{Spectrum (black points) 
obtained with the MICHELLE spectrometer at the Gemini North 
Observatory.  Also shown are the flux densities and bandpass 
coverage of 5 photometric MICHELLE filters (red) and the broad IRAS 12 
$\mu$m filter (green) with flux density that lies above the flux density of 
the 5 
MICHELLE filters.  As noted in Section 3, the little wiggles 
near 9.5 $\mu$m are due to imperfect telluric ozone cancellation 
and not to a silicate emission feature in BP Psc.}
\end{figure}

\clearpage

\begin{figure}
\plotone{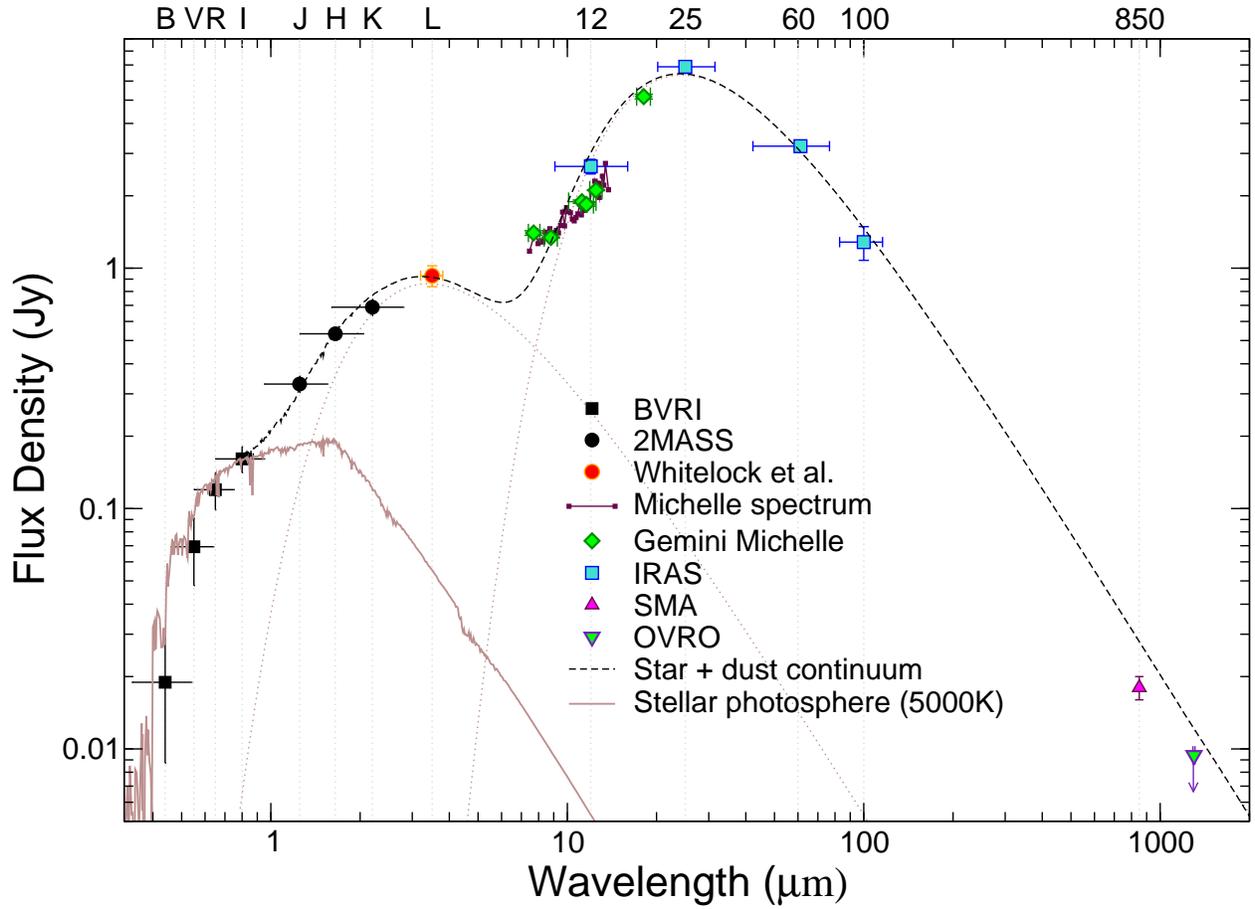}
\caption{Spectral energy distribution (SED) of BP Psc.  As 
described in Section 3, the optical points have been fitted with 
a 5000 K photosphere.  Black JHK data points are from 2MASS.  The 
orange L-band point is from Whitelock et al (1995).  The green 
points are from Table 4.  The 880 $\mu$m and 1.3 mm points are from 
Table 3.  The square blue points are from IRAS. The mid-IR spectrum is 
from 
Fig. 5.  The two dotted curves are 1500 K and 210 K blackbodies}
\end{figure}

\clearpage

\begin{figure}
\plotone{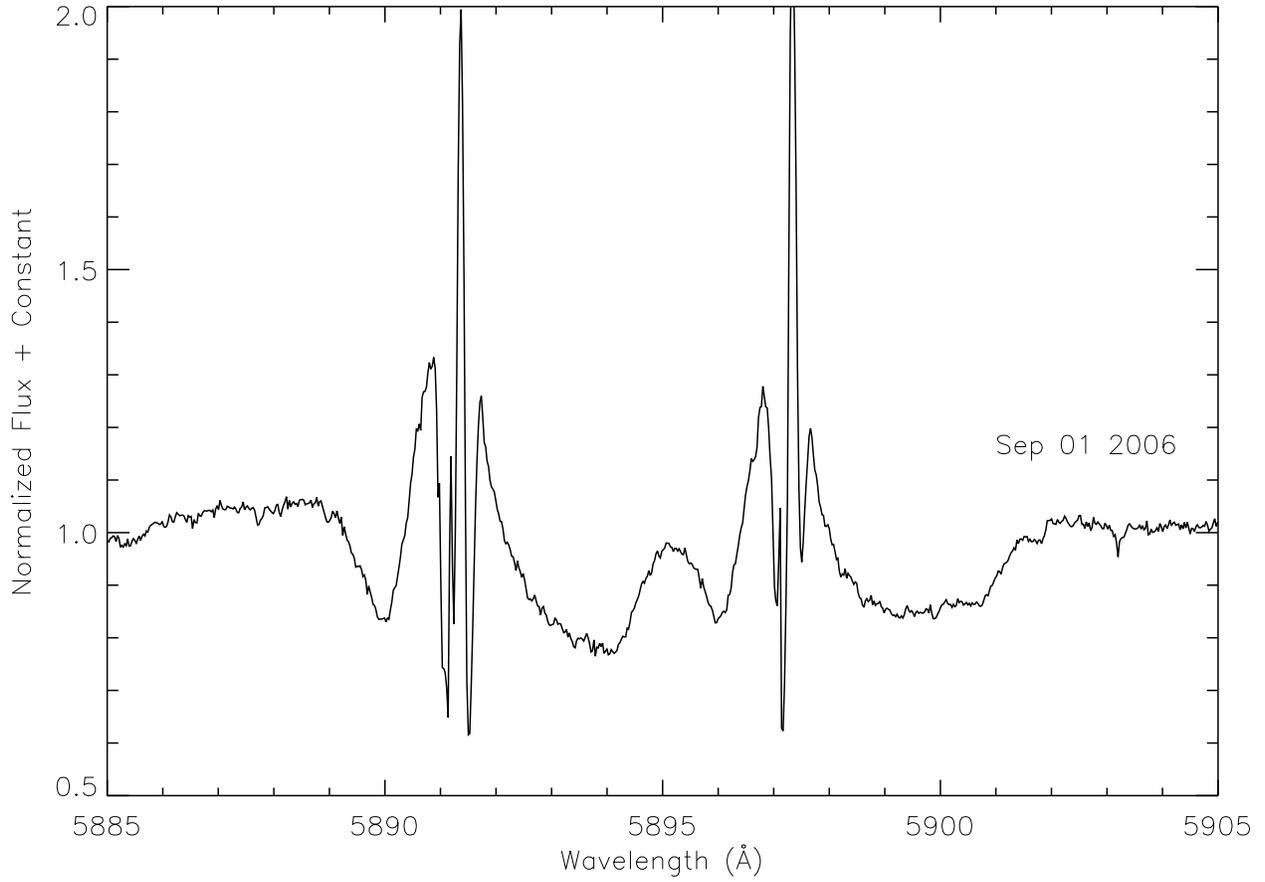}
\caption{Complex sodium D1 and D2 lines, epoch 1 Sept. 2006.  For Figures 
7-9, the wavelength scale is corrected to the heliocentric 
rest frame. One or more of the narrow absorption lines may be 
interstellar.}
\end{figure}

\clearpage

\begin{figure}
\plotone{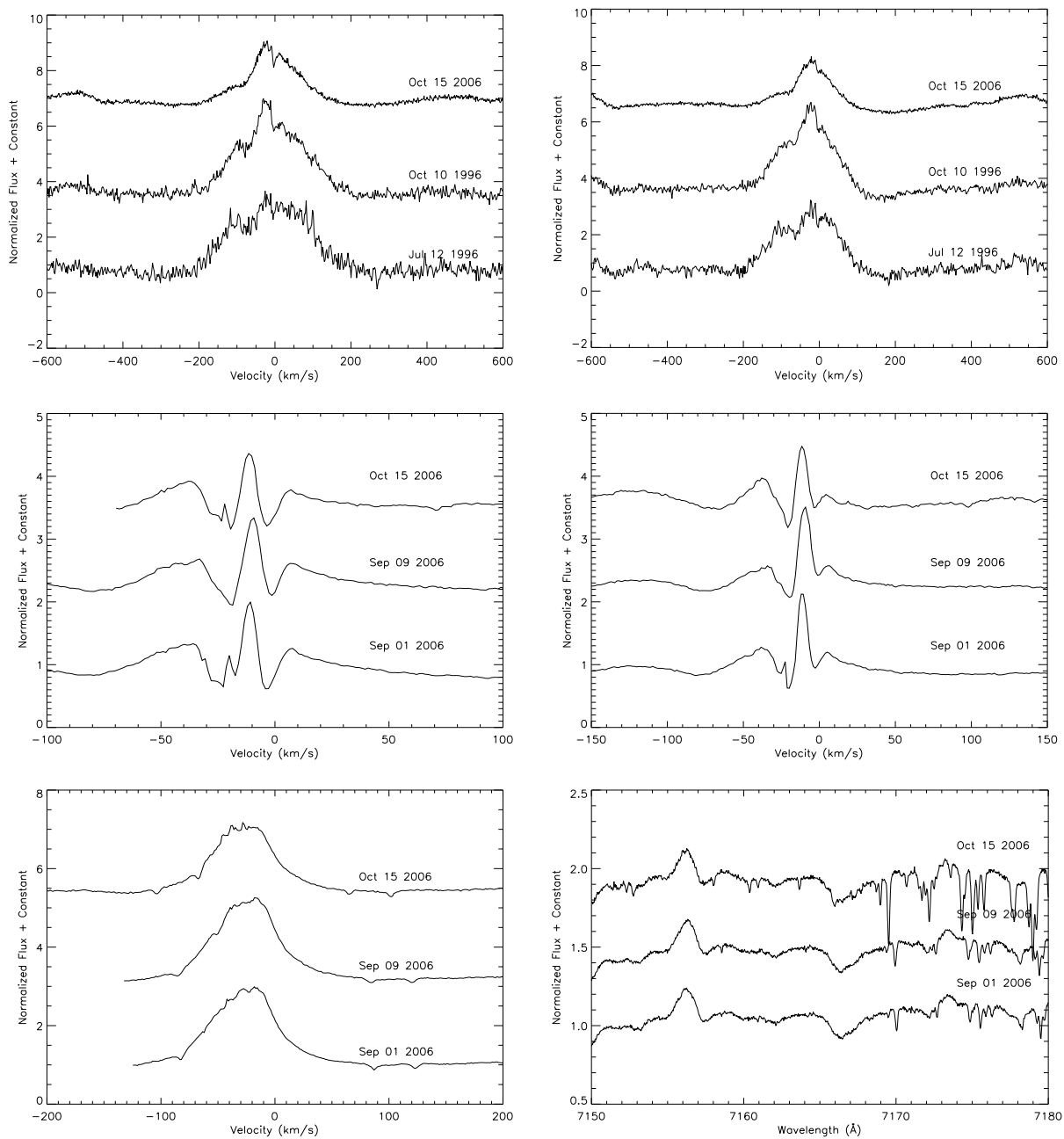}
\caption{Multiple epochs of HIRES spectra.  The indicated radial velocities 
apply to the designated transitions.  {\it Top left:} Calcium 
II K-line. {\it Top right:} Calcium II H-line. {\it Middle left:} 
Sodium D2-line. {\it Middle right:} Sodium D1-line.  {\it Lower left:}  
[OI] 6302 \AA\ line.  The associated [OI] 6365.5 \AA\ line is also 
in emission.  {\it Lower right:} [FeII] 7157.1 and 7174.0 \AA\ doublet 
emission.  Line wavelengths are from ``The Atomic Line List 2.05'' 
(www.pa.uky.edu/$\sim$peter/newpage/).  The narrow variable lines 
are telluric.}
\end{figure}

\clearpage

\begin{figure}
\epsscale{0.55}
\plotone{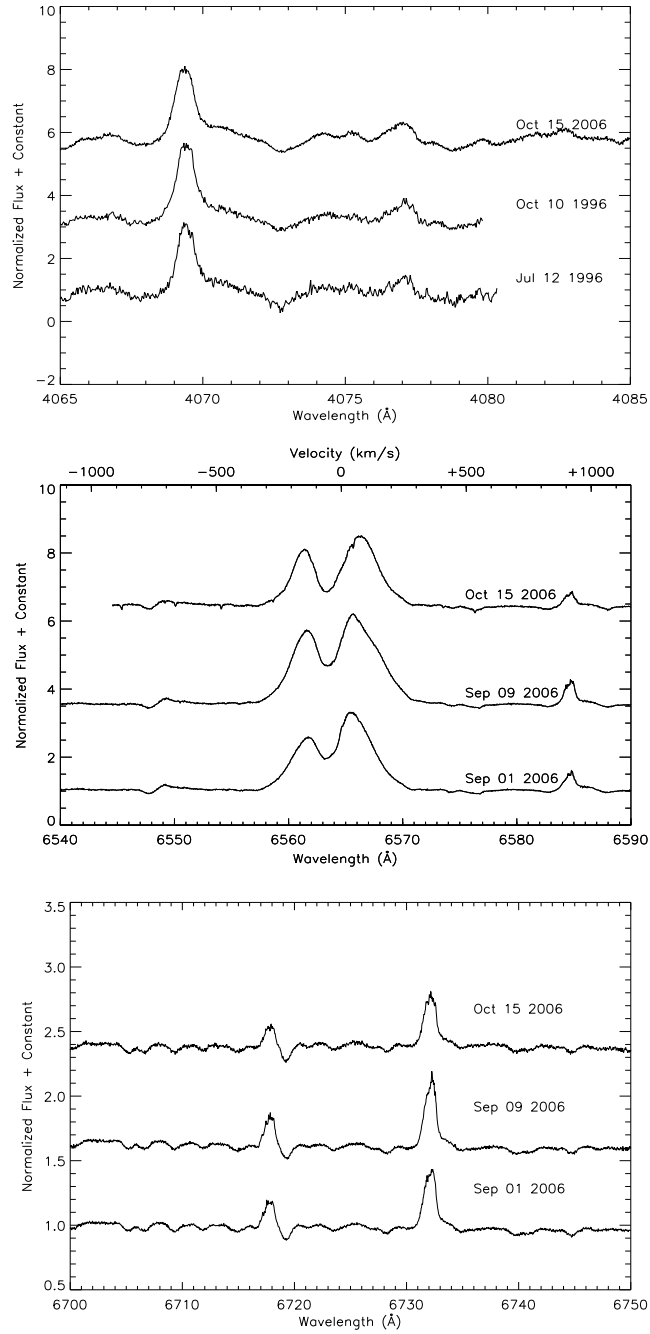}
\caption{{\it Top panel:} [SII] 
4069.7 and 4077.5 \AA\ doublet emission. {\it Middle panel:} H$\alpha$ 
and (much weaker) [NII] 6549.8 and 6585.3 \AA\ doublet emission.  The 
indicated radial velocities are based on H$\alpha$.  {\it Bottom panel:}  
Lithium I 6709.6 \AA\ in absorption and [SII] 6718.3 and 6732.7 \AA\ 
doublet emission.   For lithium, see also Fig. 15.}
\end{figure}
 
\clearpage

\begin{figure}
\epsscale{1.0}
\plotone{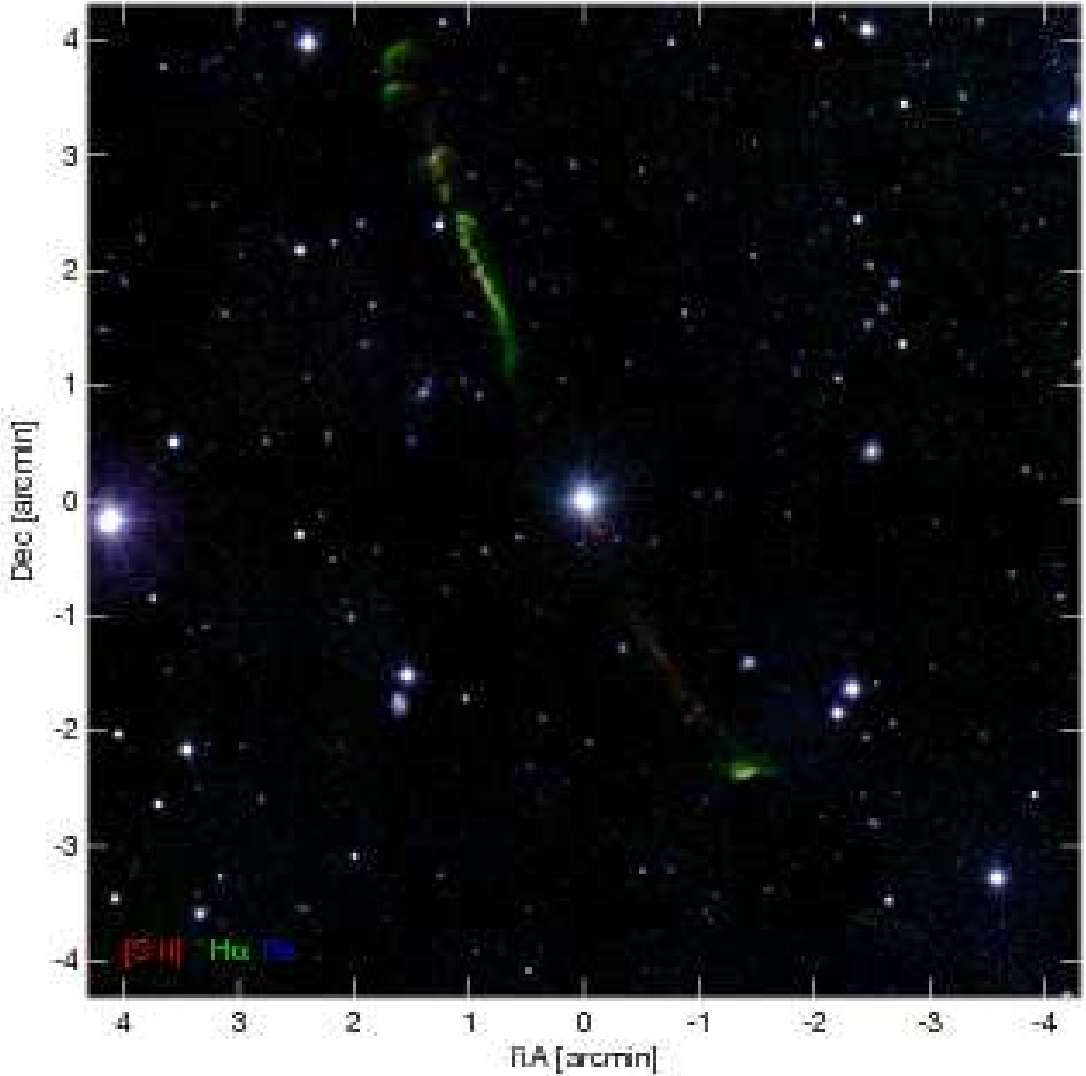}
\caption{Discovery image of outflow from BP Psc showing bipolar 
narrow emission line jets and Herbig-Haro objects, obtained at 
Lick Observatory with the 3 m Shane telescope by J. R. Graham \& 
M. D. Perrin.  The image, revealing both filamentary structure and 
bow shocks, is a composite of separate images obtained through narrow 
band H$\alpha$ and SII (6730 \AA) filters and a broadband R-filter.  
Blue is R continuum, green is H$\alpha$, and red is [SII].   The 
image has been lightly smoothed with a 1.5$\arcsec$ boxcar to highlight faint 
structure.  The nebula is most visible in H$\alpha$ emission, except 
for the two knots immediately SW of the star and a southern filament 
that appear primarily in [SII]. Note the very faint candidate HH 
object at ($-2.5$, $-3.5$) labeled as knot S6 in Table 7.}
\end{figure}

\clearpage

\begin{figure}
\epsscale{1.0}
\plotone{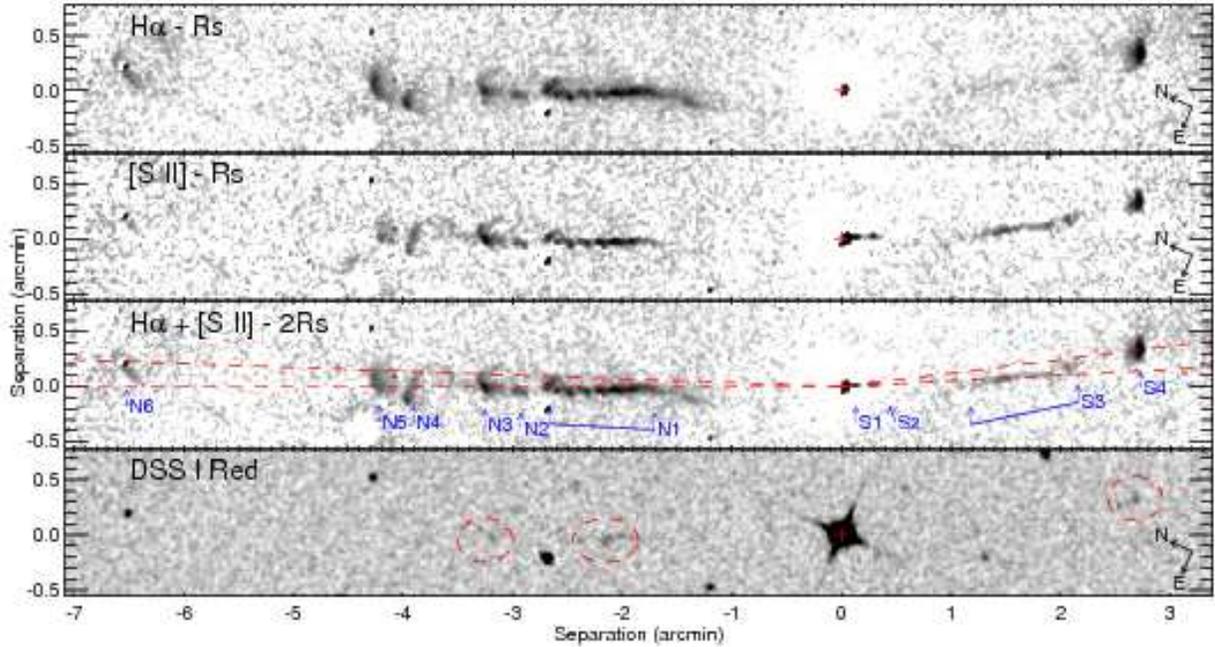}
\caption{{\it Top panel}: Continuum-subtracted H$\alpha$ image of 
the outflow from BP Psc.  The cross marks the position of BP Psc 
itself.  Due to PSF variations between the H$\alpha$ and the R images, 
a few subtraction residuals are visible at the positions of the stars 
in the field.  {\it Second panel}: Continuum-subtracted [SII] image.  
Note the two knots immediately SW (right) of BP Psc, and the much 
brighter SW filament compared to H$\alpha$.   Third panel:  Summed 
H$\alpha$ and [SII] image, with individual HH clumps labeled.  The 
dashed lines connect BP Psc with some of the brighter clumps.  
{\it Bottom panel}: For comparison a digitized Palomar Sky Survey 
red image.  Three of the brightest HH objects are visible (circled) 
in this $\sim$18 year-old plate.  Knot S4 appears to have moved about 
5$\arcsec$ between the DSS epoch and fall 2006, suggesting a proper motion 
$\sim$0.25$\arcsec$ per year (see Section 4.9).}
\end{figure}

\clearpage

\begin{figure}
\epsscale{1.0}
\plotone{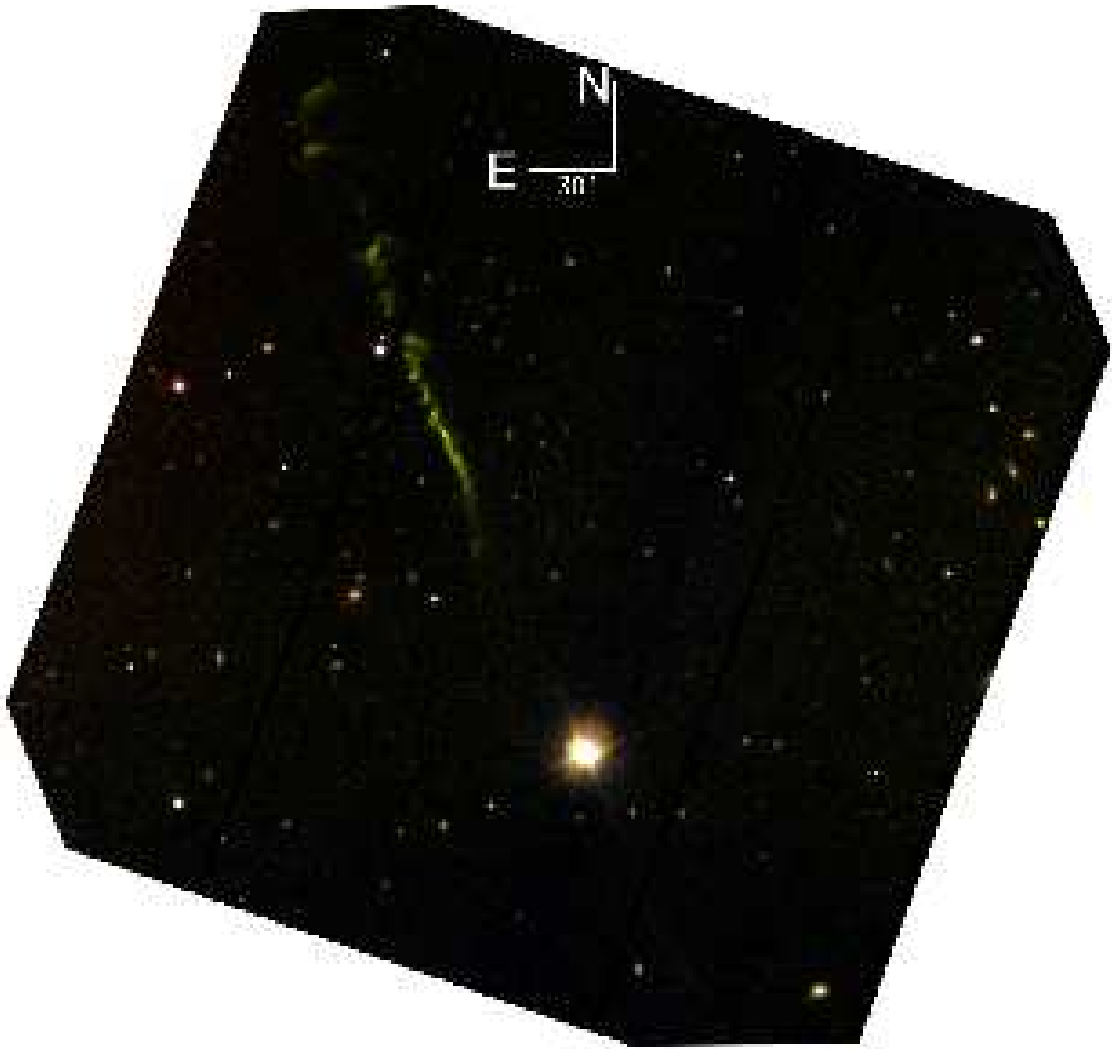}
\caption{Composite image of the NE jet obtained with the GMOS 
camera at the Gemini North Observatory.   The images were obtained at
airmass of about 2. }
\end{figure}

\clearpage

\begin{figure}   
\epsscale{1.0}   
\plotone{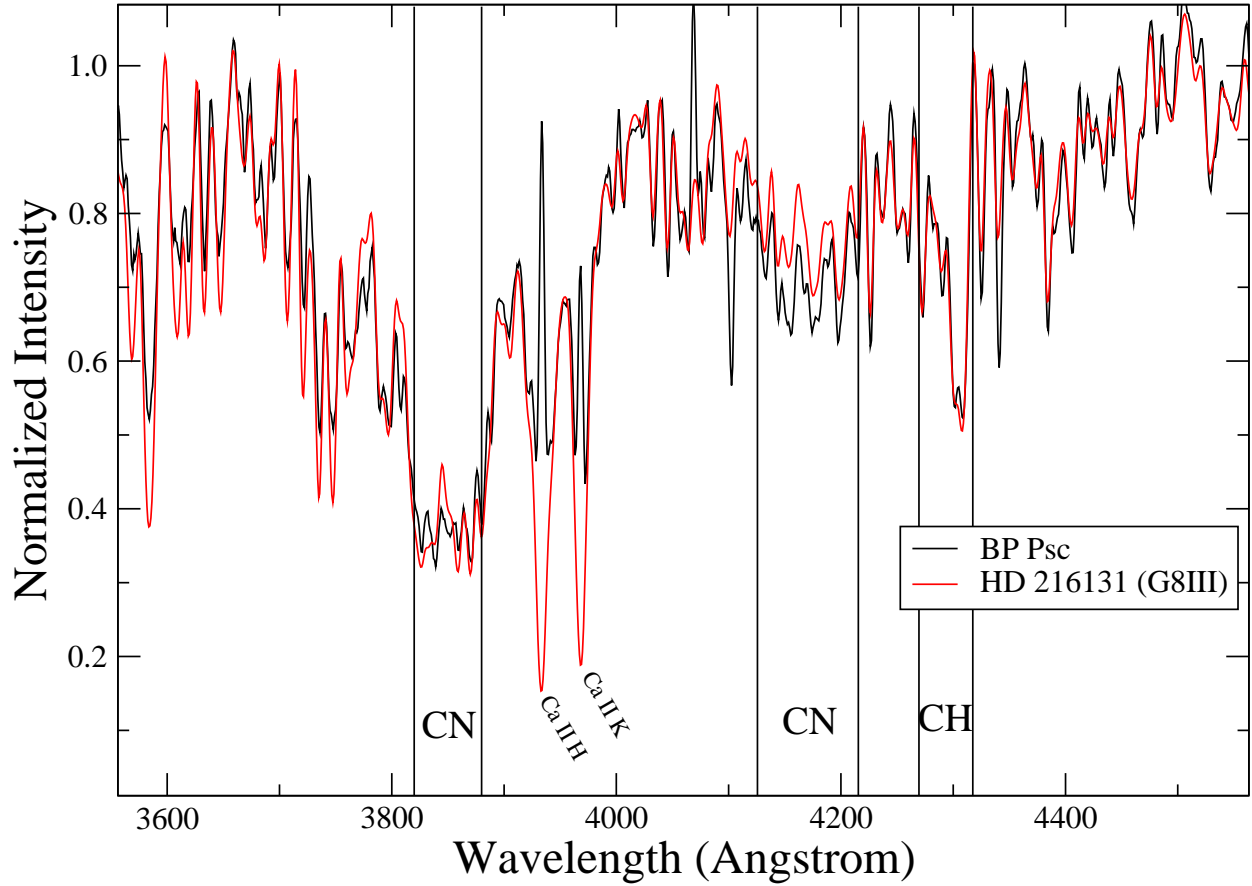}
\caption{BP Psc spectrum obtained with the Double Beam Spectrograph 
at the 2.3 m telescope at Siding Spring Observatory compared with a 
rotationally broadened spectrum of a late-G type giant from the MILES 
database (Sanchez-Blazquez et al 2006).  The CH G-band near 4300 \AA\ is 
evident as are CN bands near 3885 and 4215 \AA.}
\end{figure}

\clearpage

\begin{figure}
\epsscale{1.0}
\plotone{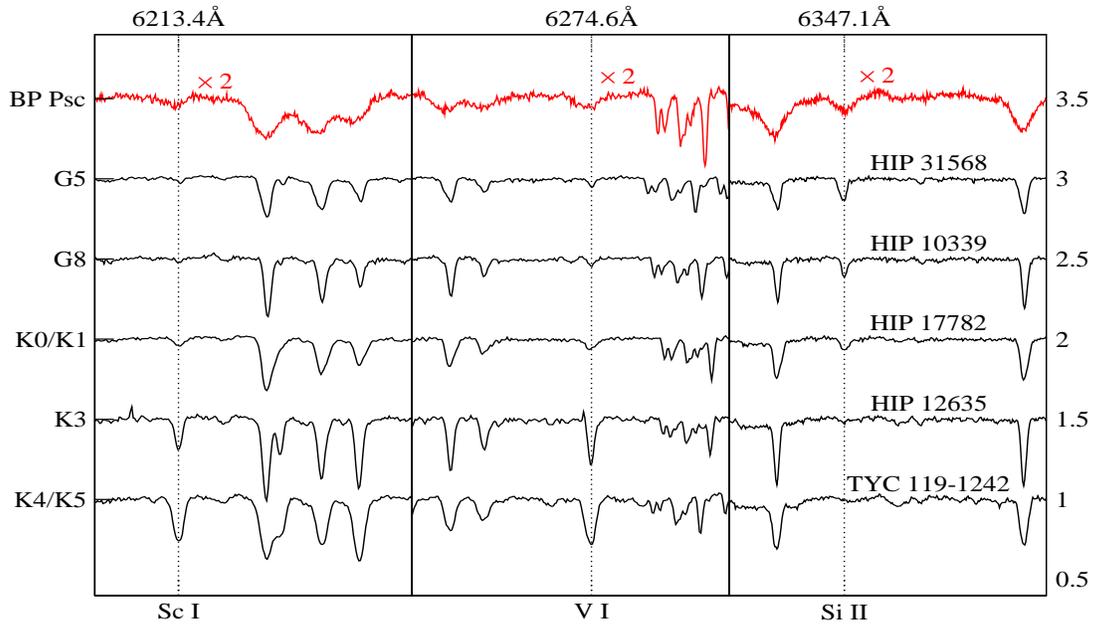}
\caption{Comparison of BP Psc HIRES spectrum with HIRES spectra of 
5 dwarf stars not known to have any surrounding dust (from Song et al 
2008).  
From comparison of the line ratios, we deduce that the spectral type 
of BP Psc, if it is a dwarf, is early K.} 
\end{figure}

\clearpage

\begin{figure}
\epsscale{1.0}
\plotone{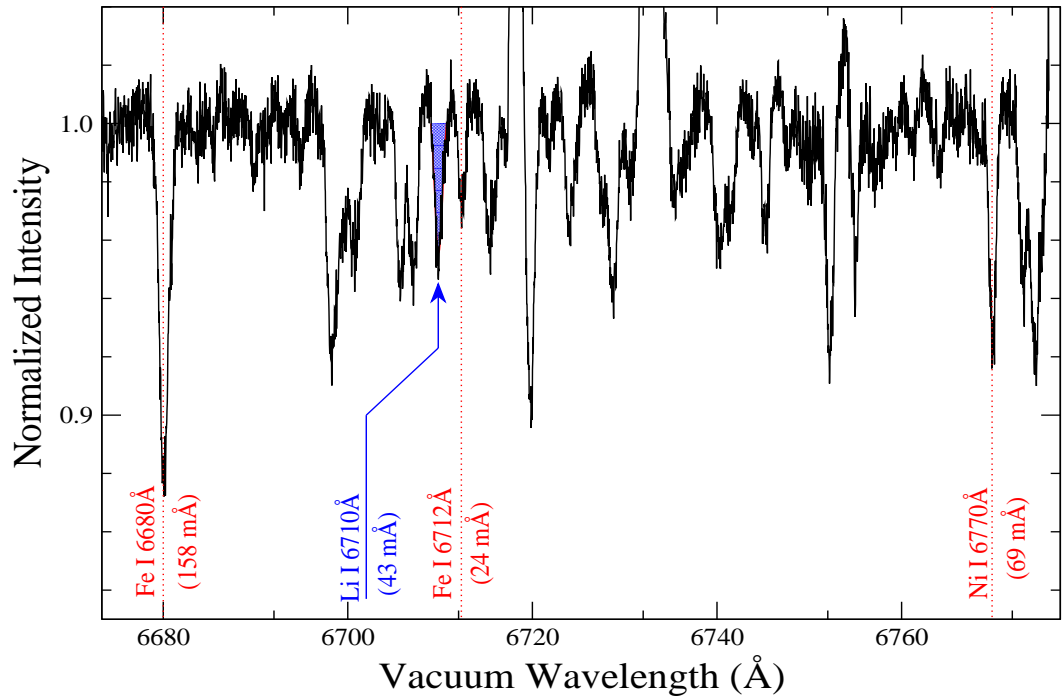}
\caption{Vicinity of the lithium 
6709.6 \AA\ line in the 9 Sept. 2006 HIRES spectrum.  As a consequence 
of veiling, the measured equivalent width (EW) of the lithium line 
should be corrected upwards as described in Section 4.3 of the text.   
Lithium EW measured on 1 Sept. and on 15 Oct. 2006 are consistent 
with the EW from 9 Sept.}
\end{figure}

\clearpage

\begin{figure}   
\epsscale{1.0}
\plotone{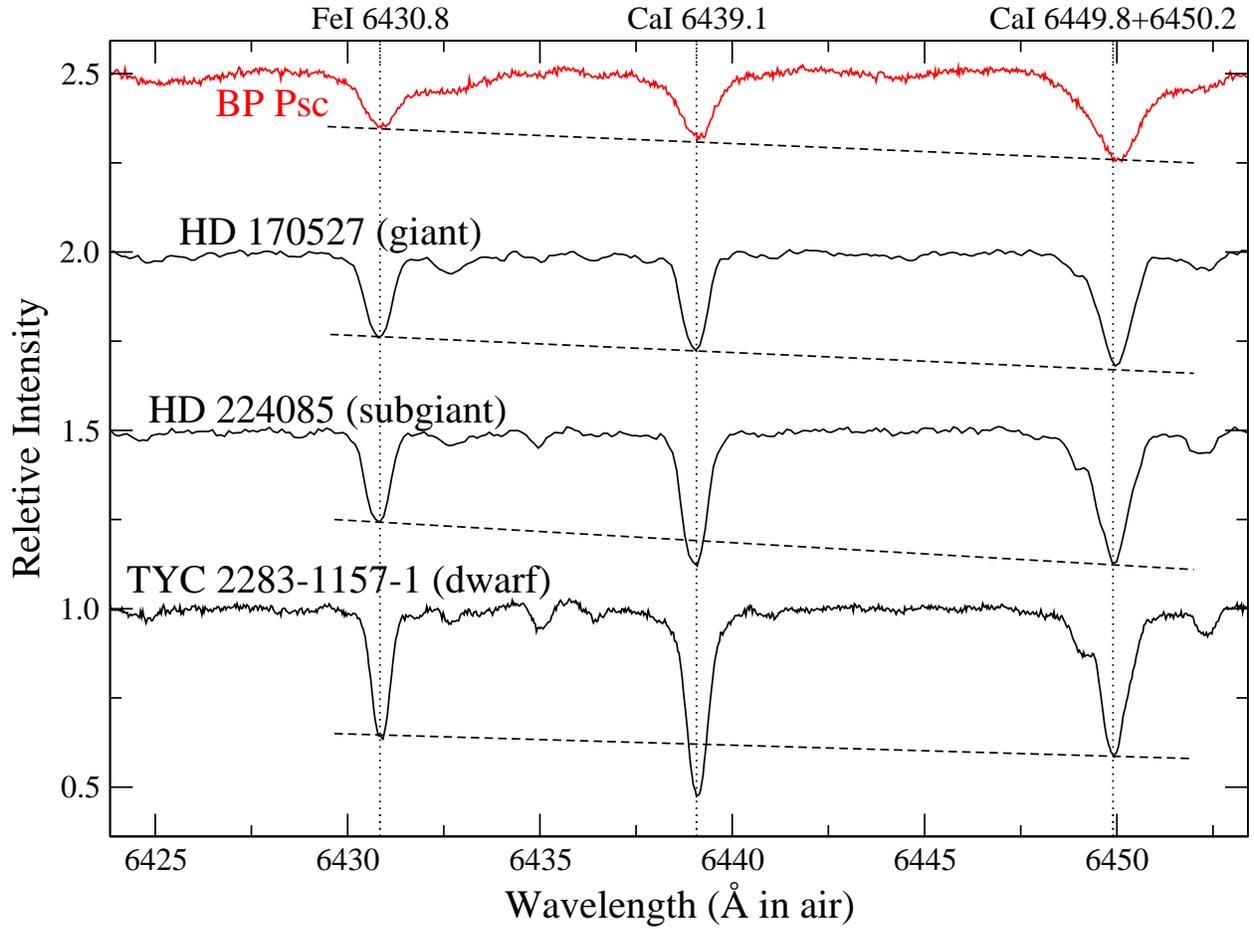}
\caption{Comparison of BP Psc HIRES spectrum with a HIRES spectrum of an
early K-type dwarf and KPNO echelle spectra (courtesy of F. Fekel) of a
G/K type giant and subgiant.}  

\end{figure}

\clearpage

\begin{figure}
\epsscale{1.0}
\plotone{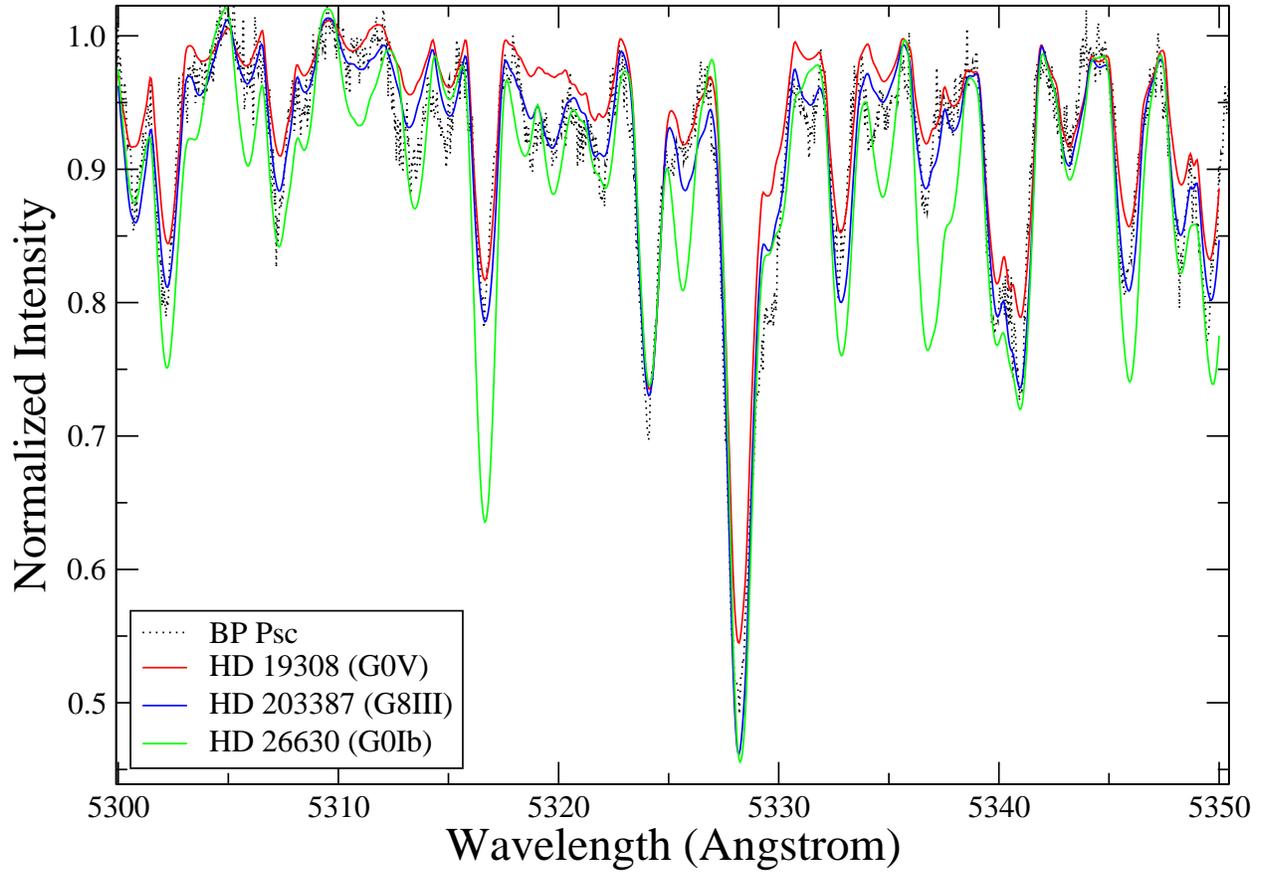}
\caption{Comparison of a BP Psc HIRES spectrum with 
spectra of a dwarf, giant 
and supergiant from the ELODIE Archive (Moultaka et al 2004) 
appropriately broadened to 
match the broad lines of BP Psc.}

\end{figure}

\clearpage

\begin{figure}
\epsscale{1.0}
\plotone{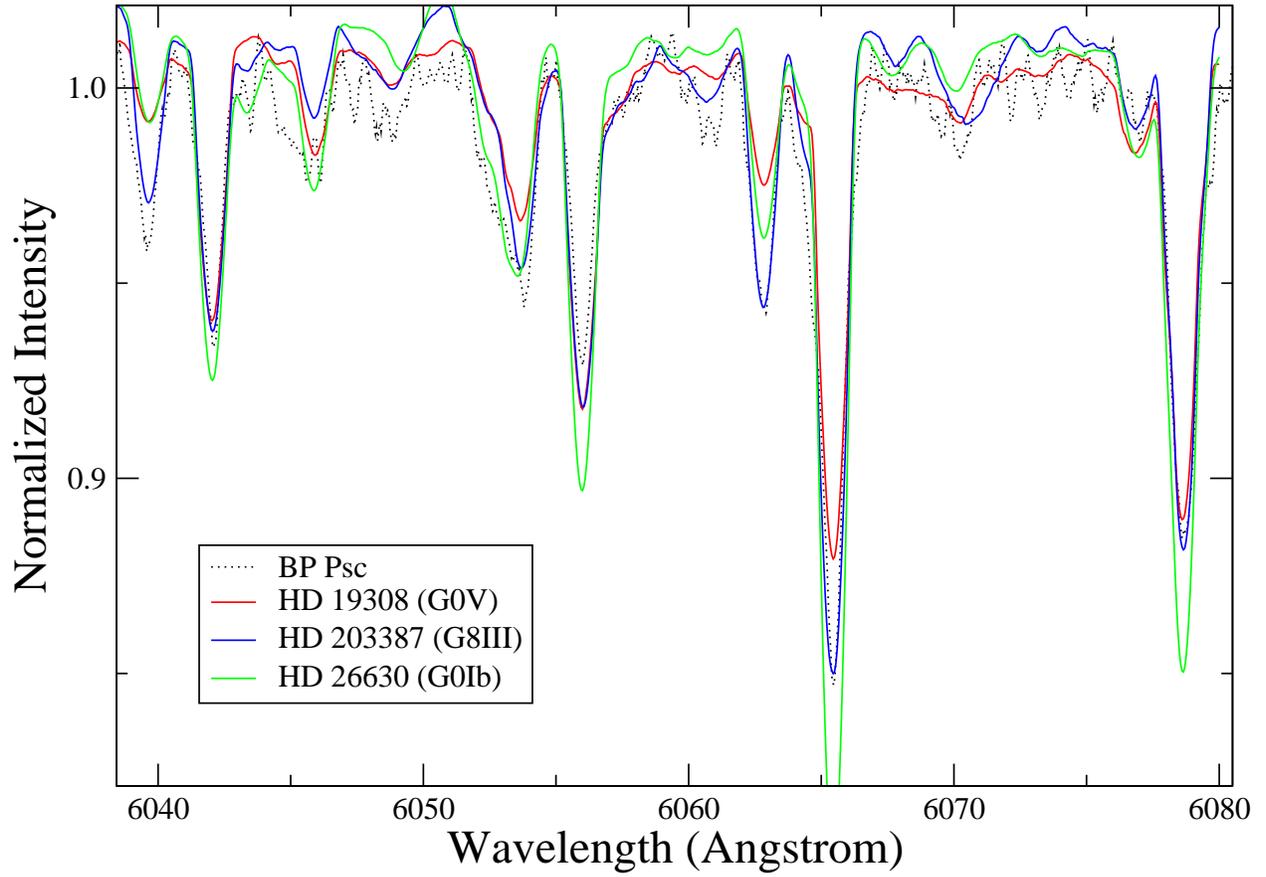}
\caption{Same as Figure 17, but for a different spectral region.}

\end{figure}

\end {document}